\documentclass[letterpaper,twocolumn,10pt]{article}
\usepackage{usenix-2020-09}

\usepackage{tikz}
\usepackage{amsmath}
\usepackage{amssymb}
\usepackage{booktabs}
\usepackage{multirow}
\usepackage{multicol}
\usepackage{bm}
\usepackage{tabularx}
\usepackage{enumerate}
\usepackage{enumitem}
\usepackage{subfigure}
\usepackage[graphicx]{realboxes}
\usepackage{svg}
\usepackage{stfloats}
\usepackage{xcolor}
\usepackage{ulem}
\usepackage{caption}
\usepackage{makecell}
\usepackage{setspace}
\usepackage{hyperref}
\usepackage{cleveref}
\usepackage{pgfplots}
\usetikzlibrary{spy}
\usetikzlibrary {decorations.fractals,spy}
\usepackage{appendix}
\newtheorem{theorem}{Theorem}[section]
\newtheorem{lemma}{Lemma}[section]

\usepackage{filecontents}

\begin{document}

\date{}

\title{\Large \bf AHSecAgg and TSKG: Lightweight Secure Aggregation for Federated Learning Without Compromise}

\author{
{\rm Siqing Zhang}\\
University of Science and \\Technology of China\\siqingzhang@mail.ustc.edu.cn
\and{\rm Yong Liao}\\University of Science and \\Technology of China\\
        yliao@ustc.edu.cn
    \and
{\rm Pengyuan Zhou \thanks{Corresponding author}}\\University of Science and \\Technology of China\\
        zpymyyn@gmail.com
}

\maketitle

\begin{abstract}
Leveraging federated learning (FL) to enable cross-domain privacy-sensitive data mining represents a vital breakthrough to accomplish privacy-preserving learning. However, attackers can infer the original user data by analyzing the uploaded intermediate parameters during the aggregation process. Therefore, secure aggregation has become a critical issue in the field of FL. Many secure aggregation protocols face the problem of high computation costs, which severely limits their applicability. To this end, we propose AHSecAgg, a lightweight secure aggregation protocol using additive homomorphic masks. AHSecAgg significantly reduces computation overhead without compromising the dropout handling capability or model accuracy. We prove the security of AHSecAgg in semi-honest and active adversary settings. In addition, in cross-silo scenarios where the group of participants is relatively fixed during each round, we propose TSKG, a lightweight Threshold Signature based masking key generation method. TSKG can generate different temporary secrets and shares for different aggregation rounds using the initial key and thus effectively eliminates the cost of secret sharing and key agreement. We prove TSKG does not sacrifice security. Extensive experiments show that AHSecAgg significantly outperforms state-of-the-art mask-based secure aggregation protocols in terms of computational efficiency, and TSKG effectively reduces the computation and communication costs for existing secure aggregation protocols.

\end{abstract}

\section{Introduction}

Distributed machine learning such as FL~\cite{mcmahan2017communication} can ensure original data remain on the local devices but requires data owners to submit local training information, which we call ``plain aggregation''. As~\cite{zhu2019deep,geiping2020inverting} indicate, attackers can infer a user's local data by utilizing only the updates submitted by users. To address this issue, numerous research efforts focus on secure aggregation~\cite{liu2022privacy}. A secure aggregation protocol needs to ensure the privacy of user data and the correctness of the server aggregation results. 

FL can be categorized into cross-device~\cite{kairouz2021advances} and cross-silo scenarios~\cite{huang2022cross}. In cross-device scenarios, users may drop out during the aggregation or have limited computation or communication resources. Therefore, for cross-device FL, we consider dropout-resilient aggregation schemes, which only involve one single server and output the same results as plain aggregation without incurring additional errors. SecAgg~\cite{bonawitz2017practical} is considered by many people to be the state-of-the-art (SOTA) secure aggregation protocol of this kind~\cite{so2021turbo,stevens2022efficient,zheng2022aggregation,liu2022sash}. SecAgg generates the masks based on the pair-wise secret keys between users. Therefore, as the number of participating users increases, its computation time becomes increasingly unbearable. In this paper, we 
propose a lightweight secure aggregation protocol, AHSecAgg, based on the additive homomorphic mask. 
Table~\ref{tab1} summarizes the comparison between AHSecAgg and two SOTA schemes.
\begin{table*}[!htbp]
	\begin{center}
		\begin{tabular}{ccccc} 
			\toprule 
			\multicolumn{2}{c}{\centering Scheme} &\makecell[c]{SecAgg~\cite{bonawitz2017practical}\\ \small(CCS 2017)}  &\makecell[c]{EffiAgg~\cite{liu2022efficient}\\ \small(TIFS 2022)}  & \makecell[c]{AHSecAgg\\ \small(Ours)}\\
			\midrule 
			\multicolumn{2}{c}{\centering The number of $(t,n)$ Shamir secret sharing per user} & $2$ & $1$ & \bm{$1$}\\
			\multicolumn{2}{c}{\centering The number of server's $(t,n)$ Shamir secret reconstruction} & $n$ & $1$ & \bm{$1$}\\
			\multicolumn{2}{c}{\centering The number of server's modular inverse operations} & $0$ & $m$ & \bm{$0$}\\
			\multirow{2}*{The number of modular exponentiations} & Server & $d(n-d)$ & $m$ & \bm{$0$}\\
            {} & User & $2n-2$ & $2m+n-1$ & \bm{$n-1$}\\
			\multirow{2}*{Computation cost} & Server & $O(mn^2)$ & $O(m\sqrt{p}+n)$ & \bm{$O(m+n)$}\\
			{} & User & $O(mn+n^2)$ & $O(m+n^2)$ & \bm{$O(m+n^2)$}\\
			\multirow{2}*{Communication cost} & Server & $O(mn+n^2)$ & $O(mn+n^2)$ & $O(mn+n^2)$\\
			{} & User & $O(m+n)$ & $O(m+n)$ & $O(m+n)$\\
			\bottomrule 
		\end{tabular}
  		\caption{Comparison between SecAgg, EffiAgg~\cite{liu2022efficient}, and AHSecAgg. $n$ indicates the total number of users, $d$ indicates the number of dropout users, $m$ indicates the length of the input vector, and the element $x$ in the input vector is from $\mathbb{Z}_p$. The comparison is under the semi-honest setting, and the results are in one aggregation. In the real world, $p \gg m \gg n > d$. When counting the number of modular exponentiations, we also include operations in key agreements and homomorphic pseudorandom generator (HPRG) expansions.}
   \label{tab1}
	\end{center}
\end{table*}

Cross-device scenarios have a dynamic participant set during each aggregation and thus users have to generate new keys for each aggregation and perform secret sharing among the current user set. But in cross-silo FL, the participant set in each aggregation is stable. We leverage this feature and propose a new Threshold Signatures based masking key generation method (TSKG) to reduce the computation cost of secure aggregation in cross-silo FL. Using TSKG, a user only needs to generate an initial key and share it during the preparation phase. He can then utilize the threshold signature algorithm to generate various temporary keys for different aggregations for masking. To reconstruct his temporary key during an aggregation, other users can generate the required temporary shares using the threshold sub-signature algorithm based on the initial shares received during the preparation phase. Moreover, since there is no need for secret sharing and transmitting shares during aggregations, users no longer need to negotiate a key for encrypted communication with each other. As such, the computation and communication benefits are significant. And we prove TSKG does not affect the capability of dropout handling and protocol security.

~\\
Our contributions can be summarized as follows: 
\begin{itemize}[leftmargin=*]  
	\item We propose AHSecAgg, a lightweight secure aggregation protocol, that outperforms the state-of-the-art mask-based schemes without compromising the dropout handling capability, accuracy, and security.
	\item We propose TSKG, a Threshold Signature based masking Key Generation method, to significantly reduce computation and communication costs for secure aggregation in cross-silo scenarios.
	\item We demonstrate the redundancy of the unknown variables in SOTA schemes and the lightweight nature of AHSecAgg via privacy analysis. 
    \item Extensive experiments show the superiority of our proposals in terms of computation overhead and applicability.
\end{itemize}

\noindent\textbf{Organization.} Section~\ref{sec:rela} provides an overview of related works on secure aggregation. In Section~\ref{section2}, we introduce the cryptographic primitives and symbols used in this paper. Section~\ref{sec3:AHS} provides a brief overview of the baseline schemes, SecAgg and EffiAgg, then introduces AHSecAgg with a privacy analysis. In Section~\ref{section4:TSKG}, we introduce TSKG, the new key generation method based on threshold signature, and provide a detailed description of its application in SecAgg. In Section~\ref{section5:sec}, we prove the security of AHSecAgg in the semi-honest and active adversary settings, and that TSKG will not sacrifice security. Section~\ref{section6:eva} presents a theoretical analysis of the complexity of AHSecAgg, the implementation details, and the evaluations. Finally, we conclude this paper in Section~\ref{section7:con}. 
\section{Related Work}
\label{sec:rela}
Secure aggregation can mainly be divided into five categories: homomorphic encryption-based, differential privacy-based, trusted execution environment-based, secure multi-party computation-based, and mask-based.

~\\
\noindent\textbf{Homomorphic encryption-based schemes} can be divided into single-key settings and multi-key settings. In the single-key settings~\cite{aono2017privacy,zhou2020privacy,zhang2020batchcrypt}, all users use the same key for encryption and decryption. However, a malicious server can create a fake user or colludes with one user to reveal the secret key. In the multi-key setting~\cite{li2021efficient,ma2022privacy}, each user has a different key. However, the nature of multi-key cryptosystems inevitably leads to a very expensive public key generation process~\cite{yin2021comprehensive}.

~\\
\noindent\textbf{Secure multi-party computation-based schemes} utilize secure multi-party computation (SMPC) protocols can achieve secure training. \cite{jayaraman2018distributed,xu2020privacy} share their local models with a set of agents. \cite{chen2021homomorphic,fang2021large} share data with multiple servers for training. These schemes always require the MPC participants to keep online during the execution and not collude~\cite{li2021survey}. 

~\\
\noindent\textbf{Trusted execution environment-based schemes.} \cite{mo2019efficient,zhao2021sear} utilize trusted hardware to ensure the confidentiality and integrity of aggregation. However, such schemes require additional hardware costs. Due to the limited memory size of trusted hardware, training within trusted hardware often has high latency~\cite{liu2022privacy}.

~\\
\noindent\textbf{Differential privacy-based schemes.} \cite{wei2020federated,hu2020personalized} use differential privacy techniques to protect private information. However, these schemes reduce the utility of user data and have a negative impact on training accuracy because of the added noise. At the same time, balancing security and data utility is also a major challenge for these approaches, which is difficult to grasp in practical applications~\cite{garfinkel2018issues}.

~\\
\noindent\textbf{Mask-based schemes.} SecAgg~\cite{bonawitz2017practical} conceals user data by adding random masks. During the aggregation process, the server cannot learn any individual user updates but can obtain the sum of updates. Compared to other types of schemes, these schemes do not involve complex cryptography computation and can be deployed on a large scale~\cite{liu2022privacy}.

There have been many works focused on improving the SOTA mask-based scheme SecAgg. NIKE~\cite{kalikinkar2018nikebased} introduces two non-colluding servers to help users negotiate pair-wise keys, but the strong security assumptions reduce its feasibility in practical applications. FastSecAgg~\cite{kadhe2020fastsecagg} reduces the cost for each user by using fast Fourier transform at the cost of weakened dropout handling capability. \cite{zheng2022aggregation} reduce computation costs by avoiding the use of secret sharing, relying on an impractical assumption that users who have submitted updates will not drop out. EffiAgg~\cite{liu2022efficient} uses an HPRG to reduce computation and communication costs by eliminating the need for negotiating pair-wise masking keys. However, it requires computing substantial discrete logarithms during unmasking, resulting in higher computation costs. SASH~\cite{liu2022sash} simplifies masking and unmasking by using a seed homomorphic pseudorandom generator (SHPRG), but the homomorphism of SHPRG used in SASH has a certain amount of error and accumulates as the number of aggregations and users increase. 

There are also some studies improving the communication structure of SecAgg. TurboAgg~\cite{so2021turbo} divides users into groups and uses a multi-group circular structure for aggregation. SecAgg+~\cite{bell2020secure} replaces the star-shaped communication topology in SecAgg with a k-regular Harray graph, reducing computation and communication costs. Improving communication structure and masking methods are orthogonal research directions. SecAgg+ uses the same masking method as SecAgg, and its advanced communication topology can also be conveniently migrated to EffiAgg or our proposal to reduce costs. Therefore, we skip further investigations into improving communication topology structure in this paper. 

Our AHSecAgg utilizes additive homomorphic masks to protect user data privacy, enabling faster masking and unmasking compared to SOTA schemes without sacrificing accuracy, security, or dropout handling capabilities. Moreover, to the best of our knowledge, TSKG is the first proposal that reuses secret sharing in secure aggregation.

\section{Cryptographic Primitives}
\label{section2}

In this section, we provide the functions and symbols utilized in the following text.
\subsection{Secret Sharing}
\label{section:ss}
In this paper, we use Shamir secret sharing~\cite{shamir1979share}, which splits a secret into $n$ shares and allows the secret to being reconstructed only when at least $t$ members collaborate. We assume the secret $sk$ and the shares $ ( sk^1, sk^2, ..., sk^n )$ are in a finite field $\mathbb{Z}_p$ for some $p$ that satisfies $p>n>t>1$. The scheme includes the following algorithms: 
\begin{itemize}[leftmargin=*]  
    \item $SSS.init(k) \rightarrow  (\mathbb{Z}_p, X )$, where $k$ is the security parameter. It randomly selects $n$ different non-zero elements $X= \{ x_{1}, x_{2}, ..., x_{n} \}$ in $\mathbb{Z}_p$ to identify the share holders $P= \{ P_{1}, P_{2}, ..., P_{n} \}$.
    \item $SSS.share(sk,t,X) \rightarrow  \{  (sk^i,x_{i} ) \}_{x_{i} \in X}$. It generates a share $sk^i$ for each participant $x_{i}$.
    \item $SSS.rec( \{ ( {sk}^{i},x_{i} ) \}_{x_{i} \in X^{'}},t)\rightarrow sk~or~\bot$. If there are at least $t$ shares collected, it outputs the reconstructed secret $sk$, otherwise it outputs an error symbol.
\end{itemize}

The security of Shamir secret sharing ensures that the shares of two different secrets among the same participants set are indistinguishable. Shamir secret sharing is additive homomorphic~\cite{goldreich1998secure}, which means that the sum of shares of different secrets can be used to reconstruct the sum of the secrets. For example, secret $sk_{1}$ and secret $sk_{2}$ are shared within the same participant set $X= \{ x_{1}, x_{2}, ..., x_{n} \}$. Each participant $x_{i}$ has ${sk}_{1}^{i}$ and ${sk}_{2}^{i}$, then $x_{i}$ can get ${sk}_{1}^{i}+{sk}_{2}^{i}$. More than $t$ participants can cooperate to reconstruct ${sk}_{1}+{sk}_{2}$ using $SSS.rec( \{ ( {sk}_{1}^{i}+{sk}_{2}^{i},x_{i} ) \}_{x_{i} \in X^{'}},t)\rightarrow {sk}_{1}+{sk}_{2}$.

\subsection{Digital Signature}
Digital signatures can ensure the authenticity and integrity of the message. We use the signature scheme that achieves security against universal forgery under chosen message attack (UF-CMA). The digital signature scheme consists of three algorithms.
\begin{itemize}[leftmargin=*]  
    \item $DS.gen(k)\rightarrow(sk,pk)$, where $k$ is the security parameter. It outputs a secret key $sk$ and a public key $pk$.
    \item $DS.sign(sk,m)\rightarrow(sig)$. It outputs a digital signature $sig$ on the message $m$. 
    \item $DS.verify( {sig,pk,m} )\rightarrow True~or~False$. It verifies whether the signature $sig$ is valid on $m$.
\end{itemize}

\subsection{Threshold Signature}
In a $(t,n)$ threshold signature scheme, the signing secret key is shared within a group of $n$ members who can generate sub-signatures using the secret shares. Only a valid subset of $t$ or more sub-signatures can generate the aggregate signature on behalf of the whole group. 

In particular, we use BLS threshold signature scheme~\cite{boneh2001short}, which has been proven to satisfy UF-CMA security. The scheme includes the following algorithms: 
\begin{itemize}[leftmargin=*]  
    \item $TS.init(k)\rightarrow param$, where $k$ is the security parameter. $\mathbb{G}_1$, $\mathbb{G}_2$ and $\mathbb{G}_T$ are multiplicative groups with the same order $p$, $g_1 \in \mathbb{G}_1$, $g_2 \in \mathbb{G}_2$. $e$ denotes a bilinear map: $\mathbb{G}_1 \times \mathbb{G}_2 \rightarrow \mathbb{G}_T$. Hash algorithm $h$: $\{ 0,1 \}^{*}\rightarrow\mathbb{G}_{1}$. It randomly selects $n$ different non-zero elements $X= \{ x_{1}, x_{2}, ..., x_{n} \}$ in $\mathbb{Z}_p$ to identify the share holders. It outputs $param = ( \mathbb{G}_{1},\mathbb{G}_{2},\mathbb{G}_{T},e,g_{1},g_{2},p,h,X)$. 
    \item $TS.trans(a)\rightarrow b$. It converts elements in $\mathbb{G}_1$ or $\mathbb{G}_2$ to elements in $\mathbb{Z}_p$ and vice versa. 
    \item $TS.gen(param,t)\rightarrow( {PK,SK,\{ ( {sk}^{i},x_{i} ) \}_{x_{i} \in X}} )$. It outputs a public key $PK$ and a secret key $SK$, and uses Shamir secret sharing to share $SK$ among $X$.
    \item $TS.sign( {sk}^{i},m,param )\rightarrow{sig}_{i}={h(m)}^{{sk}^{i}}$. $i$ uses its share to generate a sub-signature ${sig}_{i}$ on the message $m$. 
    \item $TS.rec( {\{ ( {sig}_{i}, x_{i} ) \}_{x_{i} \in X^{'}}},t)\rightarrow sig={h(m)}^{SK}~or~\bot$. If there are at least $t$ valid sub-signatures collected, it outputs the reconstructed signature $sig$, otherwise it outputs an error symbol.
    \item $TS.verify(sig, param) \rightarrow True~or~False$. It can verify whether the aggregate signature $sig$ is valid. 
\end{itemize}
\subsection{Key Agreement}
\label{section:kg}
Key agreement protocols allow two parties to securely negotiate a secret key over an insecure communication channel for further encrypted communication. Specifically, we use the Diffie-Hellman key agreement protocol~\cite{hellman1976new}, which consists of three algorithms. 
\begin{itemize}[leftmargin=*]  
    \item $DH.init(k)\rightarrow( \mathbb{G},g,q )$. It outputs a group $\mathbb{G}$ of order $q$ with generator $g$.
    \item $DH.gen( \mathbb{G},g,q )\rightarrow( x,g^{x} )$. It outputs a secret key $x$ and a public key $g^x$. 
    \item $DH.agree( {x_{i},g^{x_{j}}} )\rightarrow( g^{x_{j}})^{x_{i}}$. It outputs a shared secret key between participant $i$ and participant $j$.
\end{itemize}

We require the scheme to satisfy the Decisional Diffie-Hellman (DDH) assumption~\cite{boneh2006decision} and the Two Oracle Diffie-Hellman (2ODH) assumption~\cite{abdalla2001oracle}.

\subsection{Symmetric Authenticated Encryption}
\label{section:ae}
Symmetric authenticated encryption can ensure the confidentiality and integrity of the message, typically including three algorithms.
\begin{itemize}[leftmargin=*]  
    \item $AE.gen(k)\rightarrow(sk)$, where $k$ is the security parameter. It outputs a secret key $sk$.
    \item $AE.enc(sk,m)\rightarrow(c)$. It encrypts the message $m$ using $sk$ and outputs the ciphertext $c$.
    \item $AE.dec( {sk,c} )\rightarrow m~or~\bot$. If $sk$ is the correct key corresponding to the ciphertext $c$ and $c$ passes integrity verification, it outputs the plaintext $m$. Otherwise, it outputs an error symbol.
\end{itemize}

We need the encryption scheme to be indistinguishable under chosen plaintext attacks (IND-CPA) and ciphertext integrity (IND-CTXT)~\cite{bellare2000authenticated}.

\subsection{Public Key Infrastructure}
\label{section:pki}
Public Key Infrastructure (PKI) can be used to ensure the security and reliability of identity authentication during network communication. PKI is comprised of one or more Certificate Authorities (CA). CA is responsible for issuing digital certificates to verify the identity of users and preventing users from being defrauded or attacked during communication.

With the help of PKI, the server and each user $i$ can register their identity $( S,{PK}_{S} )$ and $( i,{PK}_{i} )$ on a bulletin board. The bulletin board only allows each party to register his own keys, and other users can use the public keys to verify signatures or encrypt messages.

\section{Lightweight Secure Aggregation}
\label{sec3:AHS}
We add random masks to the user data to ensure that the private information within the data remains inaccessible to the server. It is crucial to guarantee that the server can still correctly aggregate the sum of private data even in the presence of dropout users. We need to add different random masks to each element in a vector, otherwise the private information could be easily leaked~\cite{liu2022efficient}. We assume that the elements in input vectors are from $\mathbb{Z}_p$.

\textbf{Threat model}: Same with SecAgg and Effiagg, we consider two threat models. The first is the semi-honest model, in which the malicious party is honest but curious, meaning that the malicious party participates in the protocol honestly but will try to infer the privacy of other honest parties based on the received messages. The second is the active adversary model, in which the malicious party arbitrarily modifies or forges messages during the protocol to swindle the privacy of honest parties. Our protocol can guarantee the security of the honest parties' privacy under both of these threat models. In these models, we assume the server can be malicious and malicious users do not exceed one-third of the total users, aligned with SecAgg and EffiAgg. Note that, same with SecAgg, we do not consider preventing other attacks such as data poisoning~\cite{tolpegin2020data} or model inversion~\cite{hitaj2017deep}.

\subsection{Baselines}

We first provide a brief introduction to SecAgg and EffiAgg to help readers better understand the advantages of our proposal.

~\\
\textbf{SecAgg}. 
In SecAgg, each user generates two secret keys $s$ and $b$. The users need to use DH key agreement protocol to generate pair-wise keys between each other. Assuming that the input vector of user $i$ is $\mathbf{x_i}$, PRG is a pseudorandom generator, and $s_{i,j}$ is the pair-wise key between user $i$ and user $j$. User $i$ sends $\mathbf{y_i}$ to the server:
\begin{equation}
    \mathbf{y}_{\mathbf{i}} = \mathbf{x}_{\mathbf{i}} + \mathbf{P}\mathbf{R}\mathbf{G}\left( b_{i} \right) + {\sum_{i < j}{\mathbf{P}\mathbf{R}\mathbf{G}\left( s_{i,j} \right)}} - {\sum_{j < i}{\mathbf{P}\mathbf{R}\mathbf{G}\left( s_{j,i} \right)}}.
\end{equation}

After each user generates two secret keys $s$ and $b$, they share their secrets among all the current online users. If a user drops out, the server can collect enough shares to reconstruct the dropout user's key and then unmask correctly.

~\\
\textbf{EffiAgg}. In EffiAgg, each user only needs to generate one secret key, and there is no need to generate pair-wise keys between each pair of users. $\mathbb{G}$ is a finite cyclic group of order $q$ with generator $g$. $\mathbf{x_i}$ is the input vector of user $i$, $HPRG$ is a homomorphic pseudorandom generator, and $s_i$ is the secret key of user i. User $i$ sends $\mathbf{y_i}$ to the server:
\begin{equation}
    \mathbf{y}_{\mathbf{i}} = \mathbf{g}^{\mathbf{x}_{\mathbf{i}}} \cdot \mathbf{H}\mathbf{P}\mathbf{R}\mathbf{G}\left( s_{i} \right) = \mathbf{g}^{\mathbf{x}_{\mathbf{i}}} \cdot \mathbf{r}_{\mathbf{i}}.
\end{equation}

After unmasking, the aggregation result obtained by the server is:
\begin{equation}
    \mathbf{g}^{\mathbf{z}} = \mathbf{g}^{\sum_{u_i \in \mathcal{U}}\mathbf{x_i}}.
\end{equation}
Therefore, the server additionally needs to compute the discrete logarithm on each dimension of the result. Computing one discrete logarithm requires approximately $O(\sqrt{p})$ computation cost, where $p$ is the maximum of the value range.
\subsection{AHSecAgg}
\label{subsec:ahs}
Although EffiAgg has made significant improvements over SecAgg, the final step of computing discrete logarithms still requires considerable additional computation cost. For large networks, the computation time required for discrete logarithms can be unacceptable in practice. Next, we introduce AHSecAgg, which reduces the computation cost by skipping computing discrete logarithms.

~\\
\textbf{Additive masking method}. We assume $r \in \mathbb{Z}_p$ is a public parameter and $s \in \mathbb{Z}_p$ is the secret key randomly chosen by the user. We add the mask to a value $x\in \mathbb{Z}_p$ by: 
\begin{equation}
y=x+rs.
\end{equation}
Specifically, we define: 
\begin{equation}
E(s)=rs.
\end{equation}
When extending to masking vectors, we assume that the input vector of user $i$ is $\mathbf{x_i}=(x_i^{(1)},x_i^{(2)},…,x_i^{(m)})$, where $r$ and $E(\cdot)$ are public knowledge between the server and users. User $i$ randomly chooses a secret key, $s_i \in \mathbb{Z}_p$, to mask the vector $\mathbf{x_i}$ into $\mathbf{y_i}$ and sends it to the server:
\begin{equation}
\mathbf{y}_{\mathbf{i}} = \left( {x_{i}^{(1)} + E\left( s_{i} \right),~x_{i}^{(2)} + E\left( {E\left( s_{i} \right)} \right),\ldots} \right).
\end{equation}

\noindent\textbf{Unmasking and handling dropouts}. We can observe the additive homomorphism of E(·):
\begin{equation}
E\left( {s_{1} + s_{2}} \right) = E\left( s_{1} \right) + E\left( s_{2} \right).
\end{equation}
Utilizing this property, the server can compute the sum of the received masked vectors: 
\begin{equation}
    \begin{aligned}
            {\sum\mathbf{y}_{\mathbf{i}}} & = \left( {{\sum x_{i}^{(1)}} + {\sum{E\left( s_{i} \right)}},{\sum x_{i}^{(2)}} + {\sum{E\left( {E\left( s_{i} \right)} \right)}},\ldots} \right) \\
    & = \left( {{\sum x_{i}^{(1)}} + E\left( {\sum s_{i}} \right),{\sum x_{i}^{(2)}} + E\left( {E\left( {\sum s_{i}} \right)} \right),\ldots} \right).
    \end{aligned}
\end{equation}

After user $i$ generates the secret key $s_i$, he immediately shares $s_i$ among all the currently online users. The server can get the user list $\mathcal{U}$ of those who have submitted vectors. In the unmasking round, each online user $i$ sends the sum of the shares of users in $\mathcal{U}$ to the server:
\begin{equation}
s_{sum}^i = {\sum_{j \in \mathcal{U}}s_j^{i}},
\end{equation}
where $s_j^i$ represents the share of $s_j$ given to user $i$. As long as the server receives shares from more than $t$ users, it can reconstruct the sum of the secret keys of $\mathcal{U}$:
\begin{equation}
\left. SSS.rec\left( \left\{ \left( {s_{sum}^i,i} \right) \right\}_{i \in \mathcal{U}^{'}},t \right)\rightarrow{\sum_{j \in \mathcal{U}}s_{j}} \right. .
\end{equation}
As the server can directly reconstruct the sum of the secret keys of all the online users, there is no need for special treatment for dropouts. In this process, no single user's secret key is leaked, so the server cannot unmask any single received vector.

Based on the knowledge of $\sum_{}{s_i}$, $r$, and $E(\cdot)$, the server can compute:
\begin{equation}
\mathbf{R} = \left( {E\left( {\sum s_{i}} \right),E\left( {E\left( {\sum s_{i}} \right)} \right),\ldots} \right).
\end{equation}
Then the server can compute the sum of the input vectors:
\begin{equation}
    {\sum\mathbf{x}_{\mathbf{i}}} = {\sum\mathbf{y}_{\mathbf{i}}} - \mathbf{R}.
\end{equation}
As such, the server can aggregate the same result as plain aggregation without incurring additional errors.

\subsection{Privacy Analysis}
\label{section:priana}
\textbf{Other schemes}. We assume $\mathbf{x} = ( x^{(1)},x^{(2)},\ldots,x^{(m)} )$ is an input vector, $\mathbf{s} = ( s^{(1)},s^{(2)},\ldots,s^{(m)} )$ is the mask generated using a pseudorandom generator. The masking method of SecAgg and other similar schemes can be summarized as follows:
\begin{equation}
    \begin{aligned}
        \mathbf{y} & = \left( {x^{(1)} + s^{(1)},~x^{(2)} + s^{(2)},~\ldots,~x^{(m)} + s^{(m)}} \right) \\
        & = \left( {y^{(1)},~y^{(2)},~\ldots,~y^{(m)}} \right).
    \end{aligned}
\end{equation}

The adversary knows $\mathbf{y}$, and gets equations as follows, which we call Mask Equations:
\begin{equation}
    \mathbf{ME_{others}}: \left\{ \begin{matrix}
{x^{(1)} + s^{(1)} = y^{(1)}} \\
{x^{(2)} + s^{(2)} = y^{(2)}} \\
{x^{(3)} + s^{(3)} = y^{(3)}} \\
{\cdot \cdot \cdot} \\
{x^{(m)} + s^{(m)} = y^{(m)}} \\
\end{matrix} \right. .
\end{equation}
The coefficient matrix of $\mathbf{ME_{others}}$ is:
\begin{equation}
    \mathbf{C}_{\mathbf{o}\mathbf{t}\mathbf{h}\mathbf{e}\mathbf{r}\mathbf{s}} = \begin{pmatrix}
\begin{smallmatrix}
s^{(1)} & & & & & x^{(1)} & & & & \\
 & s^{(2)} & & & & & x^{(2)} & & & \\
 & & s^{(3)} & & & & & x^{(3)} & & \\
 & & & {\cdot \cdot \cdot} & & & & & {\cdot \cdot \cdot} & \\
 & & & & s^{(m)} & & & & & x^{(m)} \\
\end{smallmatrix}
\end{pmatrix}.
\end{equation}The augmented matrix of $\mathbf{ME_{others}}$ is:
\begin{equation}
    \mathbf{A}_{\mathbf{others}} = \begin{pmatrix}\begin{smallmatrix}
s^{(1)} & & & & & x^{(1)} & & & & & y^{(1)} \\
& s^{(2)} & & & & & x^{(2)} & & & & y^{(2)} \\
 & & s^{(3)} & & & & & x^{(3)} & & & y^{(3)} \\
 & & & {\cdot \cdot \cdot} & & & & & {\cdot \cdot \cdot} & & {\cdot \cdot \cdot} \\
 & & & & s^{(m)} & & & & & x^{(m)} & y^{(m)} \\
\end{smallmatrix}
\end{pmatrix}.
\end{equation}

The adversary wants to get $( x^{(1)},x^{(2)},\ldots,x^{(m)} )$, but doesn't know $( s^{(1)},s^{(2)},\ldots,s^{(m)} )$. Therefore, the number of unknown variables in $\mathbf{ME_{others}}$ is $2m$. We use $rank(\cdot)$ to represent the rank of a matrix. Then, we can get: 
\begin{equation}
    rank\left( \mathbf{C_{others}} \right) = rank\left( \mathbf{A_{others}} \right) = m < 2m.
\end{equation}
$\mathbf{ME_{others}}$ has infinite solutions, so the adversary cannot solve for the unique $( x^{(1)},x^{(2)},\ldots,x^{(m)} )$ of the user. However, there is redundancy in the unknown variables of $\mathbf{ME_{others}}$ because the number of unknown variables only needs to be greater than the rank of $\mathbf{C_{others}}$ and $\mathbf{A_{others}}$ in order to ensure that the adversary cannot solve for $x$.

~\\
\textbf{AHSecAgg scheme}. We assume $s$ is a user's secret key for masking, $\mathbf{x} = ( x^{(1)},x^{(2)},\ldots,x^{(m)} )$ is the input vector, and adversary knows $r$, $E(\cdot)$, and $\mathbf{y} = (y^{(1)},y^{(2)},\ldots,y^{(m)} )$ submitted by the user. So, the adversary can get the mask equations:
\begin{equation}
    \mathbf{ME_{ours}}:
\left\{ \begin{matrix}
{x^{(1)} + r^{1}s = y^{(1)}} \\
{x^{(2)} + r^{2}s = y^{(2)}} \\
{x^{(3)} + r^{3}s = y^{(3)}} \\
{\cdot \cdot \cdot} \\
{x^{(m)} + r^{m}s = y^{(m)}} \\
\end{matrix} \right. .
\end{equation}The coefficient matrix of $\mathbf{ME_{ours}}$ is: 
\begin{equation}
    \mathbf{C}_{\mathbf{o}\mathbf{u}\mathbf{r}\mathbf{s}} = \begin{pmatrix}
\begin{smallmatrix}
{r^{1}s} & x^{(1)} & & & & \\
{r^{2}s} & & x^{(2)} & & & \\
{r^{3}s} & & & x^{(3)} & & \\
{\cdot \cdot \cdot} & & & & {\cdot \cdot \cdot} & \\
{r^{m}s} & & & & & x^{(m)} \\
\end{smallmatrix}
\end{pmatrix}.
\end{equation}The augmented matrix of $\mathbf{ME_{ours}}$ is:
\begin{equation}
    \mathbf{A}_{\mathbf{o}\mathbf{u}\mathbf{r}\mathbf{s}} = \begin{pmatrix}
\begin{smallmatrix}
{r^{1}s} & x^{(1)} & & & & & y^{(1)} \\
{r^{2}s} & & x^{(2)} & & & & y^{(2)} \\
{r^{3}s} & & & x^{(3)} & & & y^{(3)} \\
{\cdot \cdot \cdot} & & & & {\cdot \cdot \cdot} & & {\cdot \cdot \cdot} \\
{r^{m}s} & & & & & x^{(m)} & y^{(m)} \\
\end{smallmatrix}
\end{pmatrix}.
\end{equation}Then we convert $\mathbf{A_{ours}}$ to a row echelon form using Gaussian elimination:
\begin{equation}
    \begin{pmatrix}
\begin{smallmatrix}
{rs} & x^{(1)} & & & & & y^{(1)} \\
 & {- rx^{(1)}} & x^{(2)} & & & & {- ry^{(1)} + y^{(2)}} \\
 & & {- rx^{(2)}} & x^{(3)} & & & {- ry^{(2)} + y^{(3)}} \\
 & & & {\cdot \cdot \cdot} & {\cdot \cdot \cdot} & & {\cdot \cdot \cdot} \\
 & & & & {- rx^{(m - 1)}} & x^{(m)} & {- ry^{(m - 1)} + y^{(m)}} \\
\end{smallmatrix}
\end{pmatrix}.
\end{equation}

The adversary wants to get $( x^{(1)},x^{(2)},\ldots,x^{(m)} )$, but doesn't know $s$. Therefore, the number of unknown variables in $\mathbf{ME_{ours}}$ is $m+1$. Then, we can get: 
\begin{equation}
    rank\left( \mathbf{C_{ours}} \right) = rank\left( \mathbf{A_{ours}} \right) = m < m + 1.
\end{equation}
Therefore, AHSecAgg can protect input vectors with the same level of privacy as other schemes while reducing the unnecessary overhead of masking by eliminating redundant unknown variables.

\subsection{Details of AHSecAgg}
Same with SecAgg and Effiagg, in our scheme, the server and users are synchronized via secure communication channels. If the server does not receive a message from a user within a certain time period, it considers that the user has dropped out. If the protocol does not abort during execution, it outputs the sum of input vectors of the remaining user set. Figure~\ref{fig1} describes the details of AHSecAgg and the underlined steps in red are designed to guarantee security in the active adversary setting.
\begin{figure*}[!htbp]
\begin{center}
\begin{tabular}{|p{17.2cm}|}  
\hline
\multirow{2}* {\centerline{AHSecAgg: Additive Homomorphic Mask based Secure Aggregation Protocol} } \\ 
{}\\
\textbf{Participants}: A server and user set $\mathcal{U}$, there is a secure communication channel between each user and the server.\\
\textbf{Public Inputs}: the number of users $n$, the threshold $t$, a finite field $\mathbb{Z}_p$, a group $\mathbb{G}$ of order $q$ with generator $g$, masking parameters $r$ and $E(\cdot)$ \textcolor{red}{\uline{, users' public keys for signatures $\{ sig_i^{pk}\}_{i \in \mathcal{U}}$ and the server's public key for signatures $sig_S^{pk}$}}.\\
\textbf{Private Inputs}: vectors $\{ \mathbf{x_i}\}_{i \in \mathcal{U}}$\textcolor{red}{\uline{, users' secret keys for signatures $\{ sig_i^{sk}\}_{i \in \mathcal{U}}$ and the server's secret}}\textcolor{red}{\uline{ key for signatures $sig_S^{sk}$}}.\\
$~\bullet~$\textbf{Round 0 Key Agreements}

\quad User $i$:

\quad\quad - Generates $(com_i^{sk},com_i^{pk}) \leftarrow DH.gen(\mathbb{G},g,q)$ \textcolor{red}{\uline{and computes $\sigma_i^1 \leftarrow DS.sign(sig_i^{sk},com_i^{pk})$}}.

\quad\quad - Sends $( {{com}_{i}^{pk}\textcolor{red}{\uline{\parallel \sigma_{i}^{1}}}} )$ to the server.

\quad Server:

\quad\quad - Collects at least $t$ different messages, and denotes $\mathcal{U}_1$ with this set of users. Otherwise, aborts.

\quad\quad - Broadcasts $\{ ( {i,{com}_{i}^{pk},\textcolor{red}{\uline{\sigma_{i}^{1}}} ) \}_{i \in \mathcal{U}_{1}}}$ to all users in $\mathcal{U}_1$.\\
$~\bullet~$\textbf{Round 1 Key Sharing}
    
\quad User $i$:

\quad\quad - Receives $\{ ( {i,{com}_{i}^{pk}\textcolor{red}{\uline{, \sigma_{i}^{1}}} ) \}_{i \in \mathcal{U}_{1}}}$, satisfying \textcolor{red}{\uline{$\forall j \in \mathcal{U}_{1}:DS.verify( {\sigma_{j},{sig}_{j}^{pk},{com}_{j}^{pk}} )\rightarrow True$, }}$|\mathcal{U}_1|>t$, and public keys 

\quad\quad\quad are all different. Otherwise, aborts. Then computes $(com_{i,j}) \leftarrow DH.agree(com_i^{sk},com_j^{pk})$ as the secret communication

\quad\quad\quad key with $j \in \mathcal{U}_{1}\backslash\{ i \}$.

\quad\quad - Samples a random element $s_{i} \in \mathbb{Z}_{p}$, generates $\{ ( s_i^{j},j ) \}_{j \in \mathcal{U}_{1}} \leftarrow SSS.Share( {s_i} )$. Computes $c_{i}^{j} \leftarrow AE.enc( {{com}_{i,j},s_{i}^{j}} )$\textcolor{red}{\uline{ and}}

\quad\quad\quad \textcolor{red}{\uline{ $\sigma_{i,j}^{2} \rightarrow DS.sign( {{sig}_{i}^{sk},i \parallel j \parallel c_{i}^{j}} )$}} for all $j \in \mathcal{U}_{1}\backslash\{ i \}$. Sends $\{( i \parallel j \parallel c_{i}^{j}\textcolor{red}{\uline{\parallel \sigma_{i,j}^{2}}} )\}_{j \in \mathcal{U}_{1}\backslash\{ i \} }$ to the server.

\quad Server:

\quad\quad - Collects at least $t$ different messages, and denotes $\mathcal{U}_2 \subseteq \mathcal{U}_1$ with this set of users. Otherwise, aborts.

\quad\quad - Broadcasts $\{ ( {j \parallel i \parallel c_{j}^{i}\textcolor{red}{\uline{\parallel \sigma_{j,i}^{2}}}} ) \}_{j \in \mathcal{U}_2,i\in \mathcal{U}_1,j \ne i}$ to all users in $\mathcal{U}_2$.\\
$~\bullet~$\textbf{Round 2 Masking}
    
\quad User $i$:

\quad\quad - Receives $\{ ( {j \parallel i \parallel c_{j}^{i}\textcolor{red}{\uline{\parallel \sigma_{j,i}^{2}}}} ) \}_{j \in \mathcal{U}_{2}\backslash\{ i\}}$, decrypts $s_j^i \leftarrow AE.dec(com_{i,j}, c_j^i)$\textcolor{red}{\uline{, $DS.verify( {\sigma_{j,i}^{2},{sig}_{j}^{pk},j \parallel i \parallel c_{j}^{i}} )\rightarrow True$ and }}

\quad\quad\quad \textcolor{red}{\uline{decrypts successfully for all $j \in \mathcal{U}_{2}$. Otherwise, aborts}}.

\quad\quad - Computes $\mathbf{y}_{\mathbf{i}} = ( {x_{i}^{(1)} + E( s_{i} ),x_{i}^{(2)} + E( {E( s_{i} )} ),\ldots} )$\textcolor{red}{\uline{ and $\sigma_i^3 \leftarrow DS.sign(sig_i^{sk},\mathbf{y_i})$}} and sends $(\mathbf{y_i}\textcolor{red}{\uline{\parallel \sigma_i^3}})$ to the server.

\quad Server:

\quad\quad - Collects at least $t$ different messages, and denotes $\mathcal{U}_3 \subseteq \mathcal{U}_2$ with this set of users. \textcolor{red}{\uline{For all $j \in \mathcal{U}_3$, }}

\quad\quad\quad \textcolor{red}{\uline{if $DS.verify( {\sigma_{j}^{3},{sig}_{j}^{pk},\mathbf{y}_{\mathbf{i}}} )\rightarrow False$, removes $j$ from $\mathcal{U}_3$.}}

\quad\quad - Broadcasts the list $\mathcal{U}_3$\textcolor{red}{\uline{ and $\sigma_S^4 \leftarrow DS.sign( {sig}_{s}^{sk},\mathcal{U}_{3} )$}} to all users in $\mathcal{U}_3$.\\
$~\bullet~$\textcolor{red}{\uline{\textbf{Round 3 Consistency Check}}}
    
\quad User $i$:

\quad\quad - \textcolor{red}{\uline{Receives $\mathcal{U}_3$, if $DS.verify( {\sigma_{s}^{4},{sig}_{s}^{pk},\mathcal{U}_{3}} )\rightarrow False$, aborts.}}

\quad\quad - \textcolor{red}{\uline{Sends $\sigma_{i}^{5} \leftarrow DS.sign\left( {{sig}_{i}^{sk},\mathcal{U}_{3}} \right) $ to the server.}}

\quad Server:

\quad\quad - \textcolor{red}{\uline{Collects at least $t$ different messages, and denotes $\mathcal{U}_4 \subseteq \mathcal{U}_3$ with this set of users, and sends $\{ ( {i,\sigma_{i}^{5}} ) \}_{i \in \mathcal{U}_{4}}$ to all users }}\quad\quad\quad \textcolor{red}{\uline{in $\mathcal{U}_4$.}}\\
$~\bullet~$\textbf{Round 4 Unmasking}

\quad If the protocol doesn't consist of Round 3, $\mathcal{U}_4=\mathcal{U}_3$.
    
\quad User $i$:

\quad\quad - Receives $\{ ( {i\textcolor{red}{\uline{,\sigma_{i}^{5}}}} ) \}_{i \in \mathcal{U}_{4}}$, if $|\mathcal{U}_4|<t$\textcolor{red}{\uline{ or $\exists j \in \mathcal{U}_{4}$: $DS.verify( {\sigma_{j}^{5},{sig}_{j}^{pk},|\mathcal{U}_3|})\rightarrow False$}}, aborts.

\quad\quad - Sends $s_{sum}^{i} = {\sum_{j \in \mathcal{U}_{3}}s_{j}^{i}}$\textcolor{red}{\uline{ and $\sigma_i^6 \leftarrow DS.sign( {{sig}_{i}^{sk},s_{sum}^{i}} )$}} to the server.

\quad Server:

\quad\quad - Collects at least $t$ different messages, and denotes $\mathcal{U}_5 \subseteq \mathcal{U}_4$ with this set of users. \textcolor{red}{\uline{For all $j \in \mathcal{U}_5$,}}\textcolor{red}{\uline{ if }}

\quad\quad\quad\textcolor{red}{\uline{$DS.verify( {\sigma_{j}^{6},{sig}_{j}^{pk},s_{sum}^j} )\rightarrow False$, removes $j$ from $\mathcal{U}_5$.}} Computes ${\sum_{j \in \mathcal{U}_{3}}s_{j}} \leftarrow SSS.rec( \{ ( {s_{sum}^{i},i} ) \}_{i \in \mathcal{U}_{5}},t)$ and gets 

\quad\quad\quad$\mathbf{R} = ( {E( {\sum_{j \in \mathcal{U}_{3}}s_{j}} ),E( {E( {\sum_{j \in \mathcal{U}_{3}}s_{j}} )} ),\ldots} )$.

\quad\quad - Computes and outputs ${\sum_{j \in \mathcal{U}_{3}}\mathbf{x}_{\mathbf{j}}} = {\sum_{j \in \mathcal{U}_{3}}\mathbf{y}_{\mathbf{j}}} - \mathbf{R}$.\\
{}\\
\hline
\end{tabular}
\caption{The detailed description of AHSecAgg for one aggregation. \textcolor{red}{\uline{The red and underlined parts are required in the active adversary setting, and not necessary in the semi-honest setting.}}}
\label{fig1}
\end{center}
\end{figure*}

\section{Threshold Signature based Masking Key Generation}
\label{section4:TSKG}
Following the common cross-silo setup, we assume that the entire FL participant user set, denoted as $\mathcal{U}_{0}$, is known and participates in each aggregation. In the aggregation process, there may be user dropouts, but no new users can join the task outside of $\mathcal{U}_{0}$~\cite{kairouz2021advances}. Scenarios involving dynamic participant set require additional checks on the user set. This will not be discussed in this paper.
\subsection{Reusing Secret Sharing}
Here, we explain how to reuse one secret sharing across different aggregations. Before FL starts, each user $i \in \mathcal{U}_0$ randomly selects an initial secret key $s_i \in \mathbb{Z}_p$, generates Shamir secret shares of $s_i$ and distributes the shares in $\mathcal{U}_0$. The secret sharing polynomial is $f(\cdot)$ and users are identified by $X= \{ x_{1}, x_{2}, ..., x_{n} \}$. The Lagrange coefficients used to reconstruct the key can also be computed in advance:
\begin{equation}
    l_{j} = {\prod_{o = 1,o \neq j}^{t}\frac{- x_{o}}{x_{j} - x_{o}}}~mod~p.
\end{equation}
There are three places where secret shares are required for a single aggregation.

\textbf{Masking}. Given a one-time random value $nonce$ of this aggregation, user $i$ can compute 
\begin{equation}
    {sig\_s_i} = {h(nonce)}^{s_i}.
\end{equation}$sig\_s_i$ is the temporary masking key for this aggregation. 

\textbf{Handling Dropouts}. When it is necessary to submit the share $s_i^j$ to the server, user $i$ submits a sub-signature instead of the share $s_i^j$: 
\begin{equation}
    {sig\_s_i^j} = {h(nonce)}^{s_i^j}.
\end{equation}
 
\textbf{Unmasking}. When the server has collected enough ``shares'', it needs to reconstruct the masking key for unmasking. It computes: 
\begin{equation}
    sig\_ s_i = {\prod_{j = 1}^{t}{sig\_s_i^j}^{l_{j}}}.
\end{equation}$sig\_ s_i$ is the temporary masking key of user $i$. 
$Proof~of~correctness$:
\begin{equation}
    \begin{aligned}
    {\prod_{j = 1}^{t}{{sig}\_s_i^j}^{l_{j}}} & = {\prod_{j = 1}^{t}{h(nonce)}^{s_i^jl_{j}}} = {\prod_{j = 1}^{t}{h(nonce)}^{f{(j)}l_{j}}} \\
    & = {h(nonce)}^{\sum_{j = 1}^{t}{f{(j)}l_{j}}} = {h(nonce)}^{f{(0)}} \\
    & = {h(nonce)}^{s_i}
    \end{aligned}.
\end{equation}
As we can see, the key $sig\_ s_i$ reconstructed by the server is exactly the temporary key generated by user $i$ before.

In different aggregations, users can use their initial secret keys to generate different temporary keys. The server can reconstruct the temporary keys to handle dropouts without revealing the initial secret keys. Therefore, with TSKG, each user only needs to perform secret sharing once and they can reuse the secret sharing in the manner described above.

\subsection{SecAgg with TSKG}
SecAgg generates two keys and performs two times of secret sharing in each aggregation. Adding TSKG allows the users only compute sub-signature thus reducing their computation and communication overhead of secret sharing. Furthermore, users no longer need to negotiate keys for encrypted communication. 
TSKG does not lead to changes in digital signatures and consistency check in the active adversary setting. Therefore, for simplicity, we only introduce how to combine Secagg with TSKG in the semi-honest setting, where SecAgg consists of four rounds: Advertise Keys, Share Keys, Masked Input Collection, and Unmasking. Appendix~\ref{appen:tskg} shows how to apply TSKG to SecAgg.

\section{Security Analysis}
\label{section5:sec}
We present the following lemma: if we mask private vectors using AHSecAgg, the masked vectors look uniformly random.

\begin{lemma}
\label{lemma1}
    Fix $n$, $m$, $r$, $E(\cdot)$, $p$ and $\{ \mathbf{x}_{\mathbf{i}} \}_{i \in \mathcal{U}}$, where $\forall i \in \mathcal{U}$, $\mathbf{x_i} \in \mathbb{Z}_p^{m}$. Then:
    \begin{equation}
        \begin{aligned}
            & \{ {\{ {s_{i}\overset{\$}{\leftarrow}\mathbb{Z}_{p}} \}_{i \in \mathcal{U}}:\{ ( {x_{i}^{(1)} + E( s_{i} ),x_{i}^{(2)} + E( {E( s_{i} )} ),\ldots} ) \}_{i \in \mathcal{U}}} \} \\
    & \equiv \{ {\{ {\mathbf{w}_{\mathbf{i}}\overset{\$}{\leftarrow}\mathbb{Z}_{p}^{m}} \}_{i \in \mathcal{U}}~s.t.{\sum_{i \in \mathcal{U}}\mathbf{w}_{\mathbf{i}}} = {\sum_{i \in \mathcal{U}}\mathbf{x}_{\mathbf{i}}}:\{ \mathbf{w}_{\mathbf{i}} \}_{i \in \mathcal{U}}} \}
        \end{aligned},
    \end{equation}
    where ``$\equiv$'' denotes the distributions are identical.
\end{lemma}

We can easily obtain the proof based on Section~\ref{section:priana}.
\subsection{Semi-honest Setting}
Let $k$ denote the security parameter, $S$ the server, $\mathcal{U}$ the set of participating users, and $t>1$ the secret sharing threshold. The set of semi-honest parties is denoted as $C$, which is a subset of $\mathcal{U} \cup {S}$. $\mathbf{x_\mathcal{U}}$ represents the set of input vectors from users in $\mathcal{U}$, and other symbols are the same as those in Figure~\ref{fig1}. Protocol security requires that no adversary can obtain private information from any honest party. The view of a party consists of its internal information and received messages. We define $\mathit{REAL}_{C}^{\mathcal{U},t,k}( {\mathbf{x}_{\mathcal{U}},\mathcal{U}_{1},\mathcal{U}_{2},\mathcal{U}_{3},\mathcal{U}_{4},\mathcal{U}_{5}} )$ as a random variable representing the joint view of adversaries in an actual protocol execution and $\mathit{SIM}_{C}^{\mathcal{U},t,k}( {\mathbf{x}_{\mathcal{U}},\mathcal{U}_{1},\mathcal{U}_{2},\mathcal{U}_{3},\mathcal{U}_{4},\mathcal{U}_{5}} )$ as the joint view of adversaries in a simulated protocol execution. In the semi-honest setting, $\mathcal{U}_3 = \mathcal{U}_4$. We consider two cases separately: when the server is honest, and when the server is semi-honest.

\begin{theorem}
\label{theo1}
    (Security against semi-honest users with an honest server.) For all $k,t,\mathcal{U},\mathbf{x}_{\mathcal{U}},\mathcal{U}_{1},\mathcal{U}_{2},\mathcal{U}_{3},\mathcal{U}_{4},\mathcal{U}_{5}$, and $C \subseteq \mathcal{U}$, where $|C| < t < | \mathcal{U}|$ and $\mathcal{U} \supseteq \mathcal{U}_1 \supseteq \mathcal{U}_2  \supseteq \mathcal{U}_3 \supseteq \mathcal{U}_4 \supseteq \mathcal{U}_5$, there exists a probabilistic polynomial time (PPT) simulator $SIM$ such that:
    \begin{equation}
        \begin{aligned}
            \mathit{SIM}_{C}^{\mathcal{U},t,k} & \left( \mathbf{x}_{\mathcal{U}},\mathcal{U}_{1},\mathcal{U}_2,\mathcal{U}_{3},\mathcal{U}_{4},\mathcal{U}_{5} \right)\\
    & \equiv \mathit{REAL}_{C}^{\mathcal{U},t,k}\left( \mathbf{x}_{\mathcal{U}},\mathcal{U}_1,\mathcal{U}_2,\mathcal{U}_3,\mathcal{U}_4,\mathcal{U}_5 \right)
        \end{aligned}.
    \end{equation}
\end{theorem}

$Proof. $In this setting, $C$ does not include the server $S$ and $|C|<t$, hence parties in $C$ cannot compute $SSS.rec(\cdot)$ successfully. The information that $C$ can get from the honest parties in the protocol includes only the sum of input vectors and the user identity list. Therefore, the messages sent and received by parties in $C$ during the protocol execution are independent of the inputs of honest parties. As such, we can conclude that the simulated joint views of the parties in $C$ is is indistinguishable from $REAL$.

\begin{theorem}
\label{theo2}
    (Security against semi-honest users with a semi-honest server.) For all $k,t,\mathcal{U},\mathbf{x}_{\mathcal{U}},\mathcal{U}_{1},\mathcal{U}_{2},\mathcal{U}_{3},\mathcal{U}_{4},\mathcal{U}_{5}$, and $C \subseteq \mathcal{U} \cup \{ S \}$, where $|C \setminus  \{ S\} | < t < | \mathcal{U}|$ and $\mathcal{U} \supseteq \mathcal{U}_1 \supseteq \mathcal{U}_2  \supseteq \mathcal{U}_3 \supseteq \mathcal{U}_4 \supseteq \mathcal{U}_5$, there exists a probabilistic polynomial time (PPT) simulator $SIM$ such that:
    \begin{equation}
        \begin{aligned}
            \mathit{SIM}_{C}^{\mathcal{U},t,k} & \left( \mathbf{x}_{\mathcal{U}},\mathcal{U}_{1},\mathcal{U}_2,\mathcal{U}_{3},\mathcal{U}_{4},\mathcal{U}_{5} \right)\\
    & \equiv \mathit{REAL}_{C}^{\mathcal{U},t,k}\left( \mathbf{x}_{\mathcal{U}},\mathcal{U}_1,\mathcal{U}_2,\mathcal{U}_3,\mathcal{U}_4,\mathcal{U}_5 \right)
        \end{aligned}.
    \end{equation}
\end{theorem}

$Proof.$ We use a standard hybrid argument to prove the theorem. We define a sequence of hybrid distributions ${H}_{0},{H}_{1},{H}_{2}\ldots$ to denote a series of modifications to $REAL$, which can finally get $SIM$. We prove $SIM$ and $REAL$ are indistinguishable by proving two adjacent hybrids are indistinguishable.
\begin{enumerate}[leftmargin=*]
    \item[$H_0$] In this hybrid, $SIM$ is exactly the same as $REAL$.
    \item[$H_1$] This hybrid is distributed exactly as the previous one, except we replace the communication keys obtained through $DH.agree(\cdot)$ in $\mathcal{U}_1\setminus C$ with uniformly random keys selected by $SIM$. The DDH assumption (see Section~\ref{section:kg}) guarantees the distribution of this hybrid is indistinguishable from the previous one.
    \item[$H_2$] This hybrid is distributed exactly as the previous one, except we replace $s_i^j$ sent by $i$ in $\mathcal{U}_2 \setminus C$ to $j$ with a share of $0$. In the unmasking round, honest parties still send the correct shares. Since $| {C \backslash \{ S \}} | < t$, semi-honest parties cannot reconstruct the key, and thus only the plaintext in encryption is changed. The IND-CPA security (see Section~\ref{section:ae}) of symmetric authenticated encryption guarantees the distribution of this hybrid is indistinguishable from the previous one.
    \item[$H_3$] This hybrid is distributed exactly as the previous one, except we replace the key $s_i$ of $i$ in $\mathcal{U}_3 \setminus C$ with a uniformly random element $s_i'$ in the corresponding field. The randomness of masking keys guarantees the distribution of this hybrid is indistinguishable from the previous one.
    \item[$H_4$] This hybrid is distributed exactly as the previous one, except users in $\mathcal{U}_3 \setminus C$ instead of sending: 
    \begin{equation}
        \mathbf{y}_{\mathbf{i}} = \left( {x_{i}^{(1)} + E\left( {s_{i}'} \right),x_{i}^{(2)} + E\left( {E\left( {s_{i}'} \right)} \right),\ldots} \right),
    \end{equation}they send: 
    \begin{equation}
        \mathbf{y}_{\mathbf{i}} = \left( {w_{i}^{(1)},w_{i}^{(2)},\ldots} \right),
    \end{equation}where $\{ \mathbf{w}_{\mathbf{i}} \}_{i \in \mathcal{U}_{3} \backslash C}$ are uniformly random vectors satisfying ${\sum_{i \in \mathcal{U}_{3}}\mathbf{w}_{\mathbf{i}}} = {\sum_{i \in \mathcal{U}_{3}}\mathbf{x}_{\mathbf{i}}}$. Lemma~\ref{lemma1} guarantees the distribution of this hybrid is indistinguishable from the previous one.
    \item[$H_5$] This hybrid is distributed exactly as the previous one, except in the unmasking round we replace $s_{sum}^{i} = {\sum_{j \in \mathcal{U}_{3}}s_{j}^{i}}$ sent by $i$ in $\mathcal{U}_5 \setminus C$ with ${rand}_{sum}^{i} = {\sum_{j \in \mathcal{U}_{3}}{rand}_{j}^{i}}$, where $rand_j$ is a uniformly random element in the corresponding field satisfying ${\sum_{j \in \mathcal{U}_{3}}{rand}_{j}} = {\sum_{j \in \mathcal{U}_{3}}s_{j}}$, and $rand_j^i$ is a share of $rand_j$. The security of Shamir secret sharing (see Section~\ref{section:ss}) guarantees the distribution of this hybrid is indistinguishable from the previous one.
\end{enumerate}

Therefore, the distribution of $SIM$ that has the same distribution as $H_5$ is indistinguishable from $REAL$. $SIM$ does not depend on the inputs of honest parties. $C$ can only learn about the sum of input vectors and the sum of secret keys. In addition, if too many users drop out (with the number of online users less than $t$), the protocol will abort and can still guarantee the above conclusion. The proof is completed.
\subsection{Active Adversary Setting}
In the active adversary setting, we cannot guarantee the correctness of the aggregation result because the malicious server can arbitrarily modify the result. However, we can guarantee the privacy of honest users' inputs. Compared with the semi-honest setting, we consider that adversaries can:
\begin{enumerate}[leftmargin=*]
    \item[1)] Send malformed or incorrect messages to interfere with the calculations of honest parties.
    \item[2)] Forge fake users to participate in the protocol.
    \item[3)] Attempt to forge or tamper with the messages of other parties.
    \item[4)] Attempt to send a fabricated special message.
    \item[5)] Attempt to forge some honest users' dropouts.
\end{enumerate}

Similar to SecAgg and EffiAgg, we only consider adversaries with probabilistic polynomial time computing capabilities. In response to the aforementioned attack methods, we take the following measures: 
\begin{enumerate}[leftmargin=*]
    \item[1)] Honest parties directly abort upon receiving a malformed or incorrect message at any round.
    \item[2)] By using PKI (see Section~\ref{section:pki}), we prevent the malicious server from forging fake users to participate in the protocol.
    \item[3)] By using digital signatures, we prevent malicious parties from successfully forging or tampering with the messages of other parties.
    \item[4)] By adding Round Consistency Check and increasing the limit of $t$, we make it impossible for the malicious server to obtain the keys of honest parties, regardless of how they construct fake messages in Round Masking.
    \item[5)] The simulator in the semi-honest setting knows the sum of input vectors. If a malicious server can forge some honest users' dropouts, the final aggregation result of $SIM$ will be different from the known sum, which leads to a failure. Therefore, our proof is carried out in a Random Oracle model. In this model, we add two trapdoor functions to inform $SIM$ of the sum of existing honest users' private information.
\end{enumerate}

In addition to the symbols defined in Figure~\ref{fig1}, we make some additional notations. $C \subseteq \mathcal{U} \cup \{ S \}$ denotes the set of malicious parties, $n_C$ denotes the number of malicious users satisfying $n_C <t$, and $n$ denotes the number of all users satisfying $n_{C} \leq \left\lceil \frac{n}{3} \right\rceil + 1$, same as SecAgg. $M_C$ is a function that represents what message malicious parties generate, and it is a probabilistic polynomial time algorithm. $\mathit{REAL}_{C}^{\mathcal{U},t,k}\left({\mathbf{x}_{\mathcal{U}},M_{C}} \right)$ is a random variable representing the joint view of adversaries in an actual protocol execution, and $\mathit{SIM}_{C}^{\mathcal{U},t,k}( M_{C} )$ is the joint view of adversaries in a simulated protocol execution. Messages of adversaries in $SIM$ and $REAL$ are both generated by $M_C$. As mentioned earlier, we provide $SIM$ with two trapdoor functions, which stipulate that in one execution of the protocol, $SIM$ can only access them once separately to obtain necessary information. These two ideal functions are ${Ideal}_{{\{ x_{i}\}}_{i \in \mathcal{U}\backslash C}}^{\delta}(L)$ and ${Ideal}_{{\{ s_{i}\}}_{i \in \mathcal{U}\backslash C}}^{\delta}(L)$, defined as follows:
\begin{equation}
        {Ideal}_{{\{ x_{i}\}}_{i \in \mathcal{U}\backslash C}}^{\delta}(L) = \left\{ 
        \begin{aligned}
            \sum_{i \in L}x_{i} &,~L \subseteq (\mathcal{U}\backslash C )~and~|L| \geq \delta \\
            \bot&,~otherwise \\
        \end{aligned}
        \right. ,
\end{equation}
\begin{equation}
        {Ideal}_{{\{ s_{i}\}}_{i \in \mathcal{U}\backslash C}}^{\delta}(L) = \left\{ 
        \begin{aligned}
            \sum_{i \in L}s_{i} &,~L \subseteq (\mathcal{U}\backslash C )~and~|L| \geq \delta \\
            \bot&,~otherwise \\
        \end{aligned}
        \right. .  
\end{equation}
We will prove that with only the sum of at least $\delta$ honest users' input vectors and the sum of secret keys provided, $SIM$ can simulate the joint view of malicious parties in the real execution. That is, during the real execution of the protocol, the malicious parties cannot learn any knowledge about the honest parties except for the sum of their input vectors and the sum of their secret keys.

~\\
\textbf{Threshold limitations}. 
We first restrict $t$ in different situations under the active adversary setting.

\textbf{1) Malicious users with an honest server}. In this situation, because $S$ is honest, all users receive the same user list $\mathcal{U}_3$ during Round Consistency Check. $S$ will not send different lists to different users, so $t>1$ can meet security requirements.

\textbf{2) Honest users with a malicious server}. In this situation, $S$ is malicious and can construct different user lists $\mathcal{U}_3$ and send them to different users, such as $\mathcal{U}_3$ and $\mathcal{U}_3'$. But honest users will only generate the signature of one user list according to the protocol. $| \{ {\sigma^{5}( \mathcal{U}_{3} )} \} |$ is the number of signatures of $\mathcal{U}_3'$ received by $S$, and $| \{ {\sigma^{5}( \mathcal{U}_{3}' )} \} |$ is the number of signatures of $\mathcal{U}_3'$. In Round 4 unmasking, if $| \{ {\sigma^{5}( \mathcal{U}_{3} )} \} | \ge t$ and $| \{ {\sigma^{5}( \mathcal{U}_{3}' )} \} | \ge t$, $S$ can obtain the sum of $\mathcal{U}_{3} \cap \mathcal{U}_{3}'$'s secret keys, and if $| \mathcal{U}_{3} \cap \mathcal{U}_{3}'| = 1$, $\mathcal{U}_{3} \cap \mathcal{U}_{3}' = \{ u \}$, then $S$ will obtain the secret key of user $u$ and its input vector. Therefore, in Round Consistency Check, we require that $| \{ {\sigma^{5}( \mathcal{U}_{3} )} \} | \geq t$ and $| \{ {\sigma^{5}( {\mathcal{U}_{3}'} )} \} | = n - | \{ {\sigma^{5}( \mathcal{U}_{3} )} \} | < t$, that is, $t \geq \left\lfloor \frac{n}{2} \right\rfloor + 1$.

\textbf{3) Malicious users with a malicious server}. In Round Consistency Check, malicious users can generate signatures of multiple user lists, and make $| \{ {\sigma^{5}( \mathcal{U}_{3} )} \} | \ge t$ and $| \{ {\sigma^{5}( \mathcal{U}_{3}' )} \} | \ge t$. Therefore, we need to further increase the limitation on $t$. $n_1$ is the number of honest users' signatures in $\{ {\sigma^{5}( \mathcal{U}_{3} )} \}$, and $n_2$ is the number of honest users' signatures in $\{ {\sigma^{5}( \mathcal{U}_{3}' )} \}$. To guarantee security, we require that $t$ satisfies: $t>n_1+n_C$ and $t>n_2+n_C$. Also, $n_1$, $n_2$, and $n_C$ satisfy $n_1+n_2+n_C \le n$. Therefore, we can derive that $t$ needs to satisfy $t > \frac{n + n_{C}}{2}$. Since $n_{C} \leq \left\lceil \frac{n}{3} \right\rceil + 1$, $t$ should satisfy $t \geq \left\lfloor \frac{2n}{3} \right\rfloor + 1$.

\begin{theorem}
    (Security against active malicious users with an honest server.) For all PPT algorithm $M_C,k,t,\mathcal{U},\mathbf{x}_{\mathcal{U} \backslash\mathbf{C}}$ and $C \subseteq \mathcal{U}$, where $n_C < t$, there exists a PPT simulator $SIM$ such that:
    \begin{equation}
        {\mathit{SIM}_{C}^{\mathcal{U},t,k}\left( M_{C} \right)} \equiv {\mathit{REAL}_{C}^{\mathcal{U},t,k}\left( {\mathbf{x}_{\mathcal{U}\backslash\mathbf{C}},M_{C}} \right)}.
    \end{equation}
\end{theorem}

$Proof.$ Since $S$ is honest, the proof here is similar to the proof of Theorem~\ref{theo1}. Even if malicious users attempt to forge fake messages, they cannot successfully forge valid signatures of honest users. And because $n_C < t$, $M_C$ can not reconstruct the secret keys of honest users. The only information that $C$ can obtain from honest parties is the sum of input vectors, the sum of secret keys, and user identity lists. Therefore, $M_C$ is independent of the inputs of honest parties. Hence we can conclude that the simulated joint views of the parties in $C$ are indistinguishable from $REAL$.

\begin{theorem}
    (Security against active malicious users with an active malicious server.) For all PPT algorithm $M_C,k,t,\delta = t - n_{C},\mathcal{U},\mathbf{x}_{\mathcal{U} \backslash\mathbf{C}}$ and $C \subseteq \mathcal{U} \cup \{ S \}$, where $n_C < t$, there exists a PPT simulator $SIM$ such that:
    \begin{equation}
        {\mathit{SIM}_{C}^{\mathcal{U},t,k}\left( M_{C} \right)} \equiv {\mathit{REAL}_{C}^{\mathcal{U},t,k}\left( {\mathbf{x}_{\mathcal{U}\backslash\mathbf{C}},M_{C}} \right)}.
    \end{equation}
\end{theorem}

The proof of this theorem can be found in Appendix~\ref{app:5.4}.

\subsection{Security of TSKG}
We assume $h:~h(nonce)=g_1^{nonce}$ in TSKG. Then we compute secret keys and shares: 
\begin{equation}
    {sig\_ s}_{i} = {h( {nonce} )}^{s_{i}} = g_{1}^{nonce\times s_{i}},
\end{equation}
\begin{equation}
    {sig\_ s}_{j}^{i} = {h ( {nonce} )}^{s_{j}^{i}} = g_{1}^{nonce \times s_{j}^{i}}.
\end{equation}
The DDH assumption (see Section~\ref{section:kg}) can ensure that ${sig\_ s}_{i}$ and ${sig\_ s}_{j}^{i}$ are indistinguishable from uniformly random strings. The BLS threshold signature scheme we use satisfies the UF-CMA security. Therefore, adversaries who do not know the correct secret key cannot successfully forge new valid signatures. 

We omit the security proof of the entire protocol here, as it depends on the specific aggregation protocol using TSKG. However, the above explanation can guarantee that TSKG does not compromise the security of the aggregation protocol under the semi-honest or active adversary settings.
\section{Performance Analysis}
\label{section6:eva}
In this section, we first analyze AHSecAgg theoretically, then describe the implementation details, and finally show the experimental results of AHSecAgg and TSKG. In the following text, we use $n$ to denote the number of users, $m$ the length of the input vector. Same with SecAgg, we only test schemes in the semi-honest setting, ignoring the consistency check, digital signatures, and PKI, which do not affect the asymptotic of the results.
\subsection{Theoretical Analysis}
\textbf{Computation Overhead of User}: $O(m+n^2)$. Each user needs to compute: (1) $2n$ key agreements, which takes $O(n)$ time, (2) $(t,n)$ secret shares, which takes $O(n^2)$ time, (3) mask of every dimension of the input vector, which takes $O(m)$ time, and (4) the sum of online users' keys, which takes $O(n)$ time. 

~\\
\noindent\textbf{Communication Overhead of User}: $O(m+n)$. Each user needs to send: (1) $n-1$ public keys, which is $O(n)$, (2) $n-1$ secret shares, which is $O(n)$, (3) a $m$-dimensional vector, which is $O(m)$, and (4) the sum of shares, which is $O(1)$.

~\\
\noindent\textbf{Computation Overhead of Server}: $O(m+n)$. The server needs to compute: (1) a $SSS.rec(\cdot)$, which takes $O(n)$ (Here, we use the same precomputation strategy as SecAgg), (2) the sum of received vectors, which takes $O(n)$ time, and (3) unmasking, which takes $O(m)$ time.

~\\
\noindent\textbf{Communication Overhead of Server}: $O(mn+n^2)$. The server needs to send: (1) $n(n-1)$ shares, which is $O(n^2)$, (2) $n$ $m$-dimensional vectors, which is $O(mn)$, and (3) the sums of shares, which is $O(n)$.

\subsection{Implementation} 
We implement SecAgg, EffiAgg, AHSecAgg, and TSKG using Python. Specifically, we use AES-GCM with 128-bit keys as the symmetric authenticated encryption scheme, AES in counter mode as the pseudorandom generator for SecAgg, SHA-256 to implement a homomorphic pseudorandom generator for EffiAgg, Sympy library~\cite{Meurer_SymPy_symbolic_computing_2017} to compute discrete logarithms in EffiAgg, pypbc library~\cite{pypbc} to implement bilinear operations and BLS threshold signature scheme. We run on a Linux workstation with 32GB of RAM and an Intel (R) Xeon (R) Gold 6246R CPU @ 3.40GHz. In the LAN setting, the latency is approximately 3-4 milliseconds. In the WAN setting, we run the server-side and user-side programs in instances located respectively in Xi'an, China, and California, USA (approximately 10427 kilometers apart), with a latency of approximately 230 to 240 milliseconds. In all experimental settings, the space of the elements in the input vectors is 32-bit. Since the results are identical across different aggregation rounds, we collect all results from one aggregation.
\subsection{Evaluation}
We have already proven that AHSecAgg outputs the same results as plain aggregation in Section~\ref{subsec:ahs}, so here we skip testing the accuracy but focus on the overhead of secure aggregation. Due to our advanced dropout handling method, dropout users in any round do not impact the aggregation. If users drop out in round Key Agreements, it will reduce the cost of secret sharing. If users drop out in round Key Sharing or round Unmasking, it will not impact the complexity of aggregation. The aggregation cost will only be significantly increased if users drop out in round Masking. So when simulating user dropouts, we assume that users drop out in Round Masking, that is, users fail to send masked vectors to the server. To avoid the impact of sending latency and waiting time on evaluating computation overhead, we only record the local computation time in the experiments. And the computation time of users and the server in the plain aggregation is $0$, where users send raw data vector to the server.
\begin{figure*}[!h]
\centering
\footnotesize \ref{fig3:sec} SecAgg~~~~~~~~~~~~~~~~~~~~~~~~~~~~~~\ref{fig3:eff} EffiAgg~~~~~~~~~~~~~~~~~~~~~~~~~~~~~~\ref{fig3:ahs} AHSecAgg
\begin{tabular}{cccc}
\begin{tikzpicture}[scale=0.6]
            \begin{axis}[small,font=\large,label style={font=\large},
            title = dropout rate 0\%,grid=major,legend columns = -1,
            legend entries = {SecAgg, EffiAgg, AHSecAgg},
            legend to name=fig5legend,
            legend style={line width=0pt}, 
            legend image post style={scale=0.7,line width=0.7pt},
    xmode=normal,
    ymode=log,
    xlabel=Vector Length, 
    ylabel=Time (s)
    ]
\addplot[smooth,line width=1pt,mark=square,color={rgb,255:red,230;green,111;blue,081}] plot coordinates { 
(10000,32.596328020095825)
(20000,58.94201946258545)
(30000,82.49068284034729)
(40000,96.39114928245544)
(50000,118.5490574836731)
(60000,149.3202953338623)
(70000,171.09441900253296)
(80000,179.37597346305847)
(90000,194.8497359752655)
(100000,207.52566981315613)
};
\label{fig3:sec}
\addplot[smooth,line width=1pt,mark=triangle,color={rgb,255:red,233;green,196;blue,107}] plot coordinates {

(10000,107.13595080375671)
(20000,227.7153458595276)
(30000,323.2121682167053)
(40000,434.50357842445374)
(50000,529.7730534076691)
(60000,631.0237810611725)
(70000,729.1065499782562)
(80000,837.4158577919006)
(90000,921.215047121048)
(100000,1019.368177652359)
};
\label{fig3:eff}
\addplot[smooth,line width=1pt,mark=o,color={rgb,255:red,042;green,157;blue,142}] plot coordinates {
(10000,0.5495848655700684)
(20000,1.097031831741333)
(30000,1.6458284854888916)
( 40000,2.1923606395721436)
(50000,2.7424325942993164)
( 60000,3.293750524520874)
(70000,3.8458151817321777)
(80000,4.404620885848999)
(90000,5.036761045455933)
(100000,5.549554109573364)
};
\label{fig3:ahs}
\end{axis}
\end{tikzpicture} 
&
\begin{tikzpicture}[scale=0.6]
\begin{axis}[
small,font=\large,label style={font=\large},
 title = dropout rate 10\%,
    xmode=normal,grid=major,
    ymode=log,
    xlabel=Vector Length, ylabel=Time (s),
    legend columns=3
    ]
\addplot[smooth,line width=1pt, mark=square,color={rgb,255:red,230;green,111;blue,081}] plot coordinates { 
(10000,921)
(20000,1975)
(30000,3241)
(40000,4228)
(50000,5102)
(60000,6651.3202953338623)
(70000,7888.09441900253296)
(80000,8143.37597346305847)
(90000,8667.8497359752655)
(100000,9513.52566981315613)
};
\addplot[smooth,line width=1pt,mark=triangle,color={rgb,255:red,233;green,196;blue,107}] plot coordinates {
(10000,102.13595080375671)
(20000,207.7153458595276)
(30000,311.2121682167053)
(40000,414.50357842445374)
(50000,517.7730534076691)
(60000,499.0237810611725)
(70000,579.1065499782562)
(80000,668.4158577919006)
(90000,742.215047121048)
(100000,819.368177652359)
};
\addplot[smooth,line width=1pt,mark=o,color={rgb,255:red,042;green,157;blue,142}] plot coordinates {
(10000, 0.5067780017852783)
(20000, 1.0060203075408936)
(30000,1.5042479038238525 )
(40000, 2.002143144607544)
(50000,2.522314548492 )
(60000,3.016209363937378 )
(70000, 3.5330324172973633)
(80000,4.032411336898804 )
(90000, 4.538487195968628)
(100000, 5.046799659729004)
};
\end{axis}
\end{tikzpicture}
&
\begin{tikzpicture}[scale=0.6]
            \begin{axis}[
            small,font=\large,label style={font=\large},grid=major,
            title = dropout rate 20\%,
    xmode=normal,
    ymode=log,ylabel=Time (s),
    xlabel=Vector Length
    ]
\addplot[smooth,line width=1pt,mark=square,color={rgb,255:red,230;green,111;blue,081}] plot coordinates { 
(10000,1638.596328020095825)
(20000,3177.94201946258545)
(30000,4624.49068284034729)
(40000,5953.39114928245544)
(50000,7053.5490574836731)
(60000,8811.3202953338623)
(70000,9458.09441900253296)
(80000,11893.37597346305847)
(90000,12699.8497359752655)
(100000,13422.52566981315613)
};
\addplot[smooth,line width=1pt,mark=triangle,color={rgb,255:red,233;green,196;blue,107}] plot coordinates {
(10000,81)
(20000,164)
(30000,247)
(40000,326)
(50000,403)
(60000,485)
(70000,567)
(80000,650)
(90000,730)
(100000,809)
};
\addplot[smooth,line width=1pt,mark=o,color={rgb,255:red,042;green,157;blue,142}] plot coordinates {
(10000,0.4569013118)
(20000,0.8950808048)
(30000,1.348297)
(40000,1.796180)
(50000,2.248113)
(60000,2.6842)
(70000,3.15377)
(80000,3.59273)
(90000,4.0553004)
(100000,4.49804115)
};
\end{axis}
\end{tikzpicture} 
&
\begin{tikzpicture}[scale=0.6]
\begin{axis}[
small,font=\large,label style={font=\large},grid=major,
 title = dropout rate 30\%,
    xmode=normal,
    ymode=log,ylabel=Time (s),
    xlabel=Vector Length, 
    legend pos=outer north east
    ]
\addplot[smooth,mark=square,line width=1pt,color={rgb,255:red,230;green,111;blue,081}] plot coordinates { 
(10000,2200)
(20000,3744)
(30000,5077)
(40000,6353)
(50000,7523)
(60000,8806)
(70000,10219)
(80000,12650)
(90000,13868)
(100000,15296)

};
\addplot[smooth,mark=triangle,line width=1pt,color={rgb,255:red,233;green,196;blue,107}] plot coordinates {
(10000,81)
(20000,162)
(30000,243)
(40000,324)
(50000,405)
(60000,485)
(70000,566)
(80000,678)
(90000,918)
(100000,1031)
};
\addplot[smooth,mark=o,line width=1pt,color={rgb,255:red,042;green,157;blue,142}] plot coordinates {
(10000,0.3947)
(20000,0.7735409736)
(30000,1.1683073)
(40000,1.54990720)
(50000,1.942368)
(60000,2.3229598)
(70000,2.720277)
(80000,3.099558830)
(90000,3.4904563)
(100000,3.8797097)
};

\end{axis}
\end{tikzpicture}\\

\end{tabular}
\caption{Server local computation time as the length of the input vector ($m$) increases, the number of users ($n$) is fixed to 500. The vertical axis adopts logarithmic scale.}
\label{fig3}
\end{figure*}

\begin{figure*}[!h]
\centering
\footnotesize \ref{fig4:sec} SecAgg~~~~~~~~~~~~~~~~~~~~~~~~~~~~~~\ref{fig4:eff} EffiAgg~~~~~~~~~~~~~~~~~~~~~~~~~~~~~~\ref{fig4:ahs} AHSecAgg
\begin{tabular}{cccc}
        \begin{tikzpicture}[scale=0.6]
            \begin{axis}[small,font=\large,label style={font=\large},
            title = dropout rate 0\%,grid=major,
            legend columns = -1,
            legend entries = {SecAgg, EffiAgg, AHSecAgg},
            legend to name=fig5legend,
            legend style={line width=0pt}, 
            legend image post style={scale=0.7,line width=0.7pt},
    xmode=normal,
    ymode=log,
    xlabel=Users, 
    ylabel=Time (s)
    ]
\addplot[smooth,line width=1pt,mark=square,color={rgb,255:red,230;green,111;blue,081}] plot coordinates { 
(100,26.442704 )
(200, 53.312736)
(300, 75.17854)
(400, 89.7231)
(500, 116.3687)
(600, 145.9719)
(700,175.57743 )
(800, 189.838587)
(900, 214.5761389)
(1000,236.45582 )
};\label{fig4:sec}
\addplot[smooth,line width=1pt,mark=triangle,color={rgb,255:red,233;green,196;blue,107}] plot coordinates {
(100,517.21527)
(200,515.252687)
(300,517.1501295566)
(400,517.14852)
(500,525.7467)
(600,526.9570)
(700,521.071433)
(800,523.8576)
(900,532.665194)
(1000,530.799231)
};\label{fig4:eff}
\addplot[smooth,line width=1pt,mark=o,color={rgb,255:red,042;green,157;blue,142}] plot coordinates {
(100,0.559919595)
(200,1.11176013)
(300,1.659944)
(400,2.211362)
(500, 2.761161)
(600,3.31535077)
(700,3.84636116)
(800,4.40353441)
(900,4.95792937)
(1000,5.5110645)
};
\label{fig4:ahs}
\end{axis}
\end{tikzpicture} 
&  \begin{tikzpicture}[scale=0.6]
\begin{axis}[
small,font=\large,label style={font=\large},
 title = dropout rate 10\%,
    xmode=normal,grid=major,
    ymode=log,
    xlabel=Users, ylabel=Time (s),
    legend style={at={(0.5,-0.2)},anchor=south},
    legend columns=3
    ]
\addplot[smooth,line width=1pt, mark=square,color={rgb,255:red,230;green,111;blue,081}] plot coordinates { 
(100,192.2154343)
(200,763.0959)
(300,1787.7991)
(400,3430.414499)
(500,5576.4061)
(600,8112.5864)
(700,10711.903321)
(800,12732.817272)
(900,15313.1224353)
(1000,17436.6749)
};
\addplot[smooth,line width=1pt,mark=triangle,color={rgb,255:red,233;green,196;blue,107}] plot coordinates {

(100,515.926738)
(200,517.6224920749)
(300,517.48174)
(400,520.712819814)
(500, 519.356952)
(600,520.04573)
(700,518.2466793)
(800,525.13928)
(900,534.25763)
(1000,525.7287108)
};
\addplot[smooth,line width=1pt,mark=o,color={rgb,255:red,042;green,157;blue,142}] plot coordinates {
(100,0.5146396160125732)
(200,1.0156290531158447)
(300,1.5155572891235352)
(400,2.017589807510376)
(500,2.526261568069458)
(600,3.0313162803649902)
(700,3.5400969982147217)
(800,4.047720670700073)
(900,4.548011541366577)
(1000,5.0635764598846436)
};
\end{axis}
\end{tikzpicture}
&  \begin{tikzpicture}[scale=0.6]
            \begin{axis}[
            small,font=\large,label style={font=\large},grid=major,
            title = dropout rate 20\%,
    xmode=normal,
    ymode=log,ylabel=Time (s),
    xlabel=Users
    ]
\addplot[smooth,line width=1pt,mark=square,color={rgb,255:red,230;green,111;blue,081}] plot coordinates { 
(100,310.40397095680237)
(200,1177.1093645095825)
(300,2617.505234003067)
(400,3609.830022573471)
(500,5532.953822374344)
(600,8254.154578924179)
(700,11454.3855843544)
(800,14423.353465557098)
(900,19042.062955141068)
(1000,23080.314561605453)

};
\addplot[smooth,line width=1pt,mark=triangle,color={rgb,255:red,233;green,196;blue,107}] plot coordinates {
(100,510.385534286499)
(200,507.25631308555603)
(300,509.013299703598)
(400,509.98328590393066)
(500,510.42904257774353)
(600,510.41974353790283)
(700,511.1359302997589)
(800,512.9576070308685)
(900,516.1184194087982)
(1000,514.891125202179)
};
\addplot[smooth,line width=1pt,mark=o,color={rgb,255:red,042;green,157;blue,142}] plot coordinates {
(100,0.4600491523742676)
(200,0.909369945526123)
(300,1.3581571578979492)
(400,1.801879644393921)
(500,2.256669759750366)
(600,2.7129228115081787)
(700,3.157839775085449)
(800,3.625580310821533)
(900,4.060510158538818)
(1000,4.532682180404663)
};
\end{axis}
\end{tikzpicture} 
&  \begin{tikzpicture}[scale=0.6]
\begin{axis}[
small,font=\large,label style={font=\large},grid=major,
 title = dropout rate 30\%,
    xmode=normal,
    ymode=log,
    xlabel=Users, ylabel=Time (s),
    legend pos=outer north east
    ]
\addplot[smooth,mark=square,line width=1pt,color={rgb,255:red,230;green,111;blue,081}] plot coordinates { 
(100,494.005331993103)
(200,1862.0735292434692)
(300,4148.828263044357)
(400,7306.339163541794)
(500,11421.702729225159)
(600,15305.381557703018)
(700,18319.054334640503)
(800,20696.59907245636)
(900,23540.707020044327)
(1000,26593.089638233185)

};
\addplot[smooth,mark=triangle,line width=1pt,color={rgb,255:red,233;green,196;blue,107}] plot coordinates {
(100,505.18973684310913)
(200,506.22685527801514)
(300,505.6812207698822)
(400,507.04279088974)
(500,509.53423595428467)
(600,513.4301643371582)
(700,510.3830347061157)
(800,511.0794177055359)
(900,511.28609704971313)
(1000,511.71717834472656)

};
\addplot[smooth,mark=o,line width=1pt,color={rgb,255:red,042;green,157;blue,142}] plot coordinates {
(100,0.3986961841583252)
(200,0.7809867858886719)
(300,1.167231559753418)
(400,1.5564510822296143)
(500,1.942474603652954)
(600,2.334029197692871)
(700,2.7168431282043457)
(800,3.115938663482666)
(900,3.5274717807769775)
(1000,3.891833782196045)

};

\end{axis}
\end{tikzpicture}\\
\end{tabular}
\caption{Server local computation time as the number of users ($n$) increases, the length of the input vector ($m$) is fixed to 50K. The vertical axis adopts logarithmic scale.}
\label{fig4}
\end{figure*}

\begin{figure*}[!h]
	\centering
	\begin{minipage}{0.48\textwidth}
		\centering
        \footnotesize \ref{fig5:sec} SecAgg \ref{fig5:eff} EffiAgg \ref{fig5:ahs} AHSecAgg
        \begin{tabular}{cc}
        \begin{tikzpicture}[scale=0.6]
                    \begin{axis}[
                    small,font=\large,label style={font=\large},grid=major,
            xmode=normal,
            ymode=log,
            legend columns = -1,
            legend entries = {SecAgg, EffiAgg, AHSecAgg},
            legend to name=fig5legend,
            legend style={line width=0pt}, 
            legend image post style={scale=0.7,line width=0.7pt},
            title = {Users = 500},
            xlabel=Vector Length,
            ylabel=Time (s)
            ]
        \addplot[smooth,line width=1pt,mark=square,color={rgb,255:red,230;green,111;blue,081}] plot coordinates { 
        (10000,15.52258014678955)
        (20000,30.855003118515015)
        (30000,49.01110529899597)
        (40000,61.44500160217285)
        (50000,76.97663450241089)
        (60000,96.58110356330872)
        (70000,107.7653021812439)
        (80000,125.62295937538147)
        (90000,139.1941454410553)
        (100000,162.06641221046448)
        };
        \label{fig5:sec}
        \addplot[smooth,line width=1pt,mark=triangle,color={rgb,255:red,233;green,196;blue,107}] plot coordinates {
        (10000,0.26139349937438966)
        (20000,0.3574850082397461)
        (30000,0.46189727783203127)
        (40000,0.5599199771881104)
        (50000,0.6447874546051026)
        (60000,0.7934238910675049)
        (70000,0.8286720275878906)
        (80000,0.9310102462768555)
        (90000,1.0362090110778808)
        (100000,1.1433653354644775)
        };
        \label{fig5:eff}
        \addplot[smooth,line width=1pt,mark=o,color={rgb,255:red,042;green,157;blue,142}] plot coordinates {
        (10000,0.21732909679412843)
        (20000,0.2179506540298462)
        (30000,0.2207641124725342)
        (40000,0.22501388788223267)
        (50000,0.23250765800476075)
        (60000,0.2288036346435547)
        (70000,0.23296687602996827)
        (80000,0.2380129337310791)
        (90000,0.24584335088729858)
        (100000,0.25644240379333494)
        };
        \label{fig5:ahs}
        \end{axis}
        \end{tikzpicture} 
        &
        \begin{tikzpicture}[scale=0.6]
        \begin{axis}[
        small,font=\large,label style={font=\large},grid=major,
            xmode=normal,
            ymode=log,
            xlabel=Users, 
            ylabel=Time (s),
            title = {Vector length = 50K}
            ]
        \addplot[smooth,mark=square,line width=1pt,color={rgb,255:red,230;green,111;blue,081}] plot coordinates { 
        (100,15.50516963005066)
        (200,31.700451850891113)
        (300,49.31168270111084)
        (400,67.30008625984192)
        (500,81.08201360702515)
        (600,97.78460216522217)
        (700,115.914137840271)
        (800,127.63380861282349)
        (900,146.4052062034607)
        (1000,162.02201342582703)
        
        };
        \addplot[smooth,mark=triangle,line width=1pt,color={rgb,255:red,233;green,196;blue,107}] plot coordinates {
        (100,0.5051313400268554)
        (200,0.5227049350738525)
        (300,0.5470534801483155)
        (400,0.5760888576507568)
        (500,0.6448012828826905)
        (600,0.7355977535247803)
        (700,0.8342505931854248)
        (800,0.9641932010650635)
        (900,1.1303431510925293)
        (1000,1.3509549140930175)
        };

        \addplot[smooth,mark=o,line width=1pt,color={rgb,255:red,042;green,157;blue,142}] plot coordinates {
        (100,0.03316762447357178)
        (200,0.05487847328186035)
        (300,0.08502566814422607)
        (400,0.12762267589569093)
        (500,0.18725714683532715)
        (600,0.26557216644287107)
        (700,0.38052878379821775)
        (800,0.49997868537902834)
        (900,0.6728150367736816)
        (1000,0.8813899040222168)
        };
        
        \end{axis}
        \end{tikzpicture}\\
        \end{tabular}
		\caption{User local computation time with different $m$ and $n$. The vertical axis adopts logarithmic scale.}
		\label{fig5}
	\end{minipage}\hfill
	\begin{minipage}{0.48\textwidth}
		\centering
        \footnotesize \ref{fig6:sec} SecAgg \ref{fig6:eff} EffiAgg \ref{fig6:ahs} AHSecAgg
        \begin{tabular}{cc}
        \begin{tikzpicture}[scale=0.6]
                    \begin{axis}[small,
                    font=\large,label style={font=\large},grid=major,
            title = {Users = 500},
            xlabel=Vector Length,
            legend columns = -1,
            legend entries = {SecAgg, EffiAgg, AHSecAgg},
            legend to name=fig5legend,
            legend style={line width=0pt}, 
            legend image post style={scale=0.7,line width=0.7pt},
            ylabel=Data Transfer per User (KB)
            ]
        \addplot[smooth,line width=1pt,mark=square,color={rgb,255:red,230;green,111;blue,081}] plot coordinates { 
        (10000,199)
        (20000,350)
        (30000,501)
        (40000,653)
        (50000,805)
        (60000,956)
        (70000,1107)
        (80000,1259)
        (90000,1411)
        (100000,1562)
        };
        \label{fig6:sec}
        \addplot[smooth,line width=1pt,mark=triangle,color={rgb,255:red,233;green,196;blue,107}] plot coordinates {
        (10000,172)
        (20000,324)
        (30000,475)
        (40000,626)
        (50000,778)
        (60000,932)
        (70000,1081)
        (80000,1232)
        (90000,1384)
        (100000,1536)
        };
        \label{fig6:eff}
        \addplot[smooth,line width=1pt,mark=o,color={rgb,255:red,042;green,157;blue,142}] plot coordinates {
        (10000,171)
        (20000,323)
        (30000,475)
        (40000,627)
        (50000,775)
        (60000,930)
        (70000,1081)
        (80000,1233)
        (90000,1383)
        (100000,1535)
        };
        \label{fig6:ahs}
        \end{axis}
        \end{tikzpicture} 
        &
        \begin{tikzpicture}[scale=0.6]
        \begin{axis}[small,
        font=\large,label style={font=\large},grid=major,
            xlabel=Users, 
            title = {Vector length = 50K},ylabel=Data Transfer per User (KB)
            ]
        \addplot[smooth,mark=square,line width=1pt,color={rgb,255:red,230;green,111;blue,081}] plot coordinates { 
        (100,767)
        (200,777)
        (300,786)
        (400,795)
        (500,805)
        (600,814)
        (700,823)
        (800,833)
        (900,842)
        (1000,851)
        };
        \addplot[smooth,mark=triangle,line width=1pt,color={rgb,255:red,233;green,196;blue,107}] plot coordinates {
        (100,762)
        (200,766)
        (300,769)
        (400,774)
        (500,778)
        (600,784)
        (700,786)
        (800,789)
        (900,794)
        (1000,799)
        
        };
        \addplot[smooth,mark=o,line width=1pt,color={rgb,255:red,042;green,157;blue,142}] plot coordinates {
        (100,761)
        (200,766)
        (300,770)
        (400,775)
        (500,778)
        (600,782)
        (700,787)
        (800,790)
        (900,794)
        (1000,798)
        
        };
        \end{axis}
        \end{tikzpicture} \\
        \end{tabular}
		\caption{Data transfer per user with different $n$ and $m$. Assume no dropouts.}
		\label{fig6}
	\end{minipage}
\end{figure*}

\begin{table}[!h]
    \centering
    \begin{tabular}{ccccc} 
        \toprule 
        \multirow{2}*{Scheme} & \multicolumn{4}{c}{dropout rate \%}\\
        \cline{2-5}
        {} & 0 & 10 & 20 & 30 \\
        \midrule 
         \makecell[c]{SecAgg~\cite{bonawitz2017practical}\\ \small(CCS 2017)} & 116.37 & 5102.04 & 7075.16 & 9421.70\\
         \makecell[c]{EffiAgg~\cite{liu2022efficient}\\ \small(TIFS 2022)} & 525.75 & 519.35 & 510.43 & 509.54\\
         \makecell[c]{AHSecAgg\\ \small(Ours)} & \textbf{2.76} & \textbf{2.53} & \textbf{2.26} & \textbf{1.94}\\
        \bottomrule 
    \end{tabular}
    \caption{The server computation time at different dropout rates. The vector length is fixed at 50K and the number of users is fixed at 500. The results are in seconds.}
    \label{tab2}
\end{table}

Figure~\ref{fig3} and Figure~\ref{fig4} plot the server-side local computation time under different conditions. To evaluate the performance of the three schemes in extreme cases, we simulate large-scale users and consider various dropout rates. We can observe that, as the number of users and the proportion of dropouts increase, the computation cost of SecAgg for handling dropouts increases sharply, in some cases even reaching thousands of seconds. This is because each online user needs to recover pair-wise keys and pair-wise masks with each dropout user. Although EffiAgg shows stability under different dropout rates, its computation cost for computing the discrete logarithmic is $O(m\sqrt{p})$, making it too expensive for long input vectors. On the other hand, AHSecAgg has the best performance under all dropout rates. The reason is that masking does not require participation from other users in AHSecAgg, nor does it necessitate PRG expansions or computing discrete logarithms. 
\begin{figure*}[!h]
	\centering
	\begin{minipage}{0.48\textwidth}
		\centering
        \begin{tabular}{cc}
            \begin{tikzpicture}[scale=0.6]
                        \begin{axis}[
                        small,font=\small,label style={font=\large},grid=major,
                xlabel=Vector Length,
                legend pos=north east,
                ylabel=Data expansion factor per user
                ]
            \addplot[smooth,line width=1pt,style = densely dotted,mark=o,color={rgb,255:red,151;green,141;blue,190}] plot coordinates { 
            (10000,1.0263157894736843)
            (20000,1.0165016501650166)
            (30000,1.0087912087912088)
            (40000,1.0065897858319606)
            (50000,1.0052770448548813)
            (60000,1.0055005500550056)
            (70000,1.003770028275212)
            (80000,1.0033003300330032)
            (90000,1.002932551319648)
            (100000,1.0026402640264027)
            
            };
            \addlegendentry{100 users}
            \addplot[smooth,line width=1pt,style=densely dashed,mark=triangle,color={rgb,255:red,101;green,068;blue,150}] plot coordinates {
            (10000,1.131578947368421)
            (20000,1.0693069306930694)
            (30000,1.043956043956044)
            (40000,1.0329489291598024)
            (50000,1.0263852242744063)
            (60000,1.023102310231023)
            (70000,1.0188501413760602)
            (80000,1.0173267326732673)
            (90000,1.0146627565982405)
            (100000,1.0132013201320131)
            };
            \addlegendentry{500 users}
            \addplot[smooth,line width=1pt,mark=square,color={rgb,255:red,058;green,013;blue,096}] plot coordinates {
            (10000,1.263157894736842)
            (20000,1.132013201320132)
            (30000,1.0879120879120878)
            (40000,1.0642504118616145)
            (50000,1.0527704485488127)
            (60000,1.045104510451045)
            (70000,1.0377002827521207)
            (80000,1.033828382838284)
            (90000,1.029325513196481)
            (100000,1.0264026402640265)
            };
            \addlegendentry{1000 users}
            \end{axis}
            \end{tikzpicture} 
            &
            \begin{tikzpicture}[scale=0.6]
            \begin{axis}[
            small,font=\small,label style={font=\large},grid=major,
                xlabel=Users,ylabel=Data expansion factor per user,
                legend pos=north west
                ]
            \addplot[smooth,line width=1pt,mark=square,color={rgb,255:red,058;green,013;blue,096}] plot coordinates { 
            (100,1.0263157894736843)
            (200,1.0526315789473684)
            (300,1.0789473684210527)
            (400,1.105263157894737)
            (500,1.131578947368421)
            (600,1.1578947368421053)
            (700,1.1842105263157894)
            (800,1.236842105263158)
            (900,1.263157894736842)
            (1000,1.2960526315789473)
            
            };
            \addlegendentry{10K-dimension}
            \addplot[smooth,line width=1pt,style=densely dashed,mark=triangle,color={rgb,255:red,101;green,068;blue,150}] plot coordinates {
            (100,1.0039577836411608)
            (200,1.0105540897097625)
            (300,1.0158311345646438)
            (400,1.0224274406332454)
            (500,1.0263852242744063)
            (600,1.0316622691292876)
            (700,1.0382585751978892)
            (800,1.04221635883905)
            (900,1.0474934036939314)
            (1000,1.0527704485488127)
            };
            \addlegendentry{50K-dimension}
            \addplot[smooth,line width=1pt,mark=o,style = densely dotted,color={rgb,255:red,151;green,141;blue,190}] plot coordinates {
            (100,1.0026385224274406)
            (200,1.004617414248021)
            (300,1.0072559366754616)
            (400,1.0105540897097625)
            (500,1.0131926121372032)
            (600,1.0151715039577835)
            (700,1.0178100263852243)
            (800,1.0211081794195251)
            (900,1.0237467018469657)
            (1000,1.0257255936675462)
            };
            \addlegendentry{100K-dimension}
            \end{axis}
            \end{tikzpicture}\\
        \end{tabular}
		\caption{Data expansion factor per user of AHSecAgg compared to the plain aggregation. Different lines represent different values of $m$ and $n$. Assume no dropouts.}
		\label{fig7}
	\end{minipage}\hfill
	\begin{minipage}{0.48\textwidth}
		\centering
        \begin{tabular}{cc}
            \begin{tikzpicture}[scale=0.6]
                        \begin{axis}[
                        small,font=\small,label style={font=\large},grid=major,
                xlabel=Users,
                legend pos=north west,
                ylabel=Time (s)
                ]
            \addplot[smooth,line width=1pt,mark=square,color={rgb,255:red,230;green,111;blue,081},fill={rgb,255:red,235;green,246;blue,242}] plot coordinates { 
            (200,0.17715883255004883)
            (400,0.5154719352722168)
            (600,1.1678555011749268)
            (800,2.320854425430298)
            (1000,4.13888955116272)
            (1200,6.923191785812378)
            (1400,10.91121244430542)
            (1600,16.493244171142578)
            (1800,24.033068418502808)
            (2000,33.825154066085815)
            }
            \closedcycle;
            \addlegendentry{reduce}
            \addplot[smooth,line width=1pt,mark=triangle,color={rgb,255:red,042;green,157;blue,142}] plot coordinates {
            (200,0.011658906936645508)
            (400,0.023126840591430664)
            (600,0.034825801849365234)
            (800,0.0465695858001709)
            (1000,0.05828547477722168)
            (1200,0.06899666786193848)
            (1400,0.10232734680175781)
            (1600,0.09227967262268066)
            (1800,0.10388040542602539)
            (2000,0.13308167457580566)
            };
            \addlegendentry{increase}
            \end{axis}
            \end{tikzpicture} 
            &
            \begin{tikzpicture}[scale=0.6]
            \begin{axis}[
            small,font=\small,label style={font=\large},grid=major,
                xlabel=Users,
                ylabel=Data Transfer per User (KB),
                legend pos=north west
                ]
            \addplot[smooth,line width=1pt,mark=square,color={rgb,255:red,230;green,111;blue,081}] plot coordinates { 
            (100,777)
            (200,795)
            (300,814)
            (400,833)
            (500,851)
            (600,871)
            (700,890)
            (800,909)
            (900,928)
            (1000,947)
            };
            \addlegendentry{SecAgg w/o TSKG}
            \addplot[smooth,line width=1pt,mark=triangle,color={rgb,255:red,042;green,157;blue,142}] plot coordinates {
            (100,762)
            (200,767)
            (300,770)
            (400,775)
            (500,779)
            (600,783)
            (700,788)
            (800,792)
            (900,797)
            (1000,801)
            
            };
            \addlegendentry{SecAgg w/ TSKG}
            \end{axis}
            \end{tikzpicture}
        \end{tabular}
		\caption{Left: With TSKG, the reduced and increased computation time per user of SecAgg. Right: Data transfer per user with and without TSKG. (Assume no dropouts.)}
		\label{fig8}
	\end{minipage}
\end{figure*}
\begin{table*}[!h]
    \centering
    \begin{tabular}{cccccccc} 
        \toprule 
        {} &  Users & Dropout & Key Agreement & Key Sharing & Masking & Unmasking & Total time\\
        \midrule 
         User &  500 &  -- &  3/235 &  138/369 &  20/251 &  36/269 &  197/1125 \\
         Server & 500 & 0\% & 3/232 & 4/236 & 5/234 & 2790/3015 & 2801/3718 \\
         Server & 500 & 10\% & 4/233 & 3/232 & 4/237 & 2538/2746 & 2550/3449\\
         Server & 500 & 20\% & 3/234 & 4/236 & 5/236 & 2242/2457 & 2253/3164\\
         Server & 500 & 30\% & 3/235 & 3/236 & 5/234 & 1965/2237 & 1976/2945\\
         \hline
         User & 1000 & -- & 3/235 & 801/1032 & 22/252 & 69/301 & 895/1819\\
         Server & 1000 & 0\% & 3/236 & 4/237 & 4/235 & 5631/5881 & 5643/6590\\
         Server & 1000 & 10\% & 4/236 & 4/234 & 5/238 & 5164/5260 & 5177/5968\\
         Server & 1000 & 20\% & 4/233 & 4/232 & 5/237 & 4544/4718 & 4556/5421\\
         Server & 1000 & 30\% & 4/239 & 4/231 & 4/237 & 3929/4162 & 3941/4870\\
        \bottomrule 
    \end{tabular}
    \caption{The running time of each round of AHSecAgg in the LAN/WAN, without accounting for waiting time. The simulations are in the semi-honest model. The vector length is fixed to 50K. The results are in milliseconds.}
    \label{tab3}
\end{table*}

As shown in Table~\ref{tab2}, for 30\% dropout rate, the required computation time of SecAgg is astonishingly high, reaching 9421.7 seconds, because the server needs $(500 \times 30\%) \times (500 \times (1 - 30\%)) \times 50K$ PRG expansions to remove pair-wise masks and $(500 \times (1 - 30\%)) \times 50K$ PRG expansions to remove self masks, in other words, the server needs the total PRG expansions of 2,642,500,000 for one aggregation! However, AHSecAgg does not require PRG expansions.

Figure~\ref{fig5} shows the computation cost per user. We observe that as the number of users and the length of input vectors increase, the user computation overhead in SecAgg becomes unbearable, with the magnitude even reaching 100 seconds, caused by the generation of pair-wise keys and pair-wise masks. On the other hand, as our masking method does not require a large number of modular exponentiations, AHSecAgg outperforms SecAgg and EffiAgg with the best performance.

Table~\ref{tab3} summarizes the computation time per round of each user and the server. It can be seen that AHSecAgg is very lightweight and thus applicable in practice. For example, with 500 users, 50K-dimensional input vectors, and 30\% dropout rate in LAN, AHSecAgg only needs 0.2 seconds of additional computation time per user and 2 seconds for the server, while SecAgg needs 77.8 seconds per user and 9421.7 seconds for the server.

Figure~\ref{fig6} shows the communication overhead per user. We can observe that AHSecAgg has similar communication overhead with EffiAgg when vector length is fixed, and both are lower than SecAgg. This is because they do not require users to negotiate pair-wise masking keys, thus users only need to send one key's shares in the secret sharing round and send the sum of shares instead of $n-1$ shares in the unmasking round. When the number of users is fixed, the communication overhead of the three protocols is quite similar. It's probably because the communication cost of users mainly comes from sending vectors, which is also inevitable in plain aggregation. This communication overhead is caused by the nature of FL itself. Figure~\ref{fig7} provides the proof.

Figure~\ref{fig7} shows only when the input vector is very short (e.g., 10k), our scheme incurs considerable additional communication overhead compared to plain aggregation, which rarely occurs in practice because even small models such as LeNet~\cite{lecun1998gradient} have 50K parameters. When the vector length reaches 50K, the data expansion factor per user compared to plain aggregation only reaches $1.05 \times$ even with 1000 users. As the vector length increases, the data expansion factor becomes closer to $1 \times$. Therefore, in practical applications, the additional communication overhead induced by AHSecAgg is negligible. To reduce the communication overhead of FL, future work can explore parameter compression techniques such as quantization to reduce the amount of data transmission or adopt more advanced communication structures between users.

Figure~\ref{fig8} shows that after using TSKG, users in SecAgg reduce the computational and communication costs required for $2$ secret sharing and $n-1$ key agreements.

\section{Conclusion}
\label{section7:con}
In this paper, we propose a lightweight and practical secure aggregation protocol called AHSecAgg that can be used to securely compute the sum of input vectors across multiple parties, and AHSecAgg allows some parties to drop out during the process. We prove the security of AHSecAgg in semi-honest and active adversary settings, and demonstrate the superiority of AHSecAgg through experiments. Compared with existing schemes, AHSecAgg achieves lower computation overhead and wider applicability. In addition, for secure aggregation in cross-silo scenarios, we design a Threshold Signature based masking key generation method (TSKG), which can avoid secret sharing during aggregations and reduce overhead.

\normalem
\bibliographystyle{plain}
\bibliography{reference}

\appendix
\section{Appendix}
\subsection{Details of SecAgg with TSKG}
\label{appen:tskg}
Figure~\ref{fig2} describes the details of SecAgg with TSKG.
\begin{figure*}[h]
\begin{center}
\begin{tabular}{|p{17.2cm}|}  
\hline
\multirow{2}*{\centerline {The detailed description of SecAgg with TSKG for one aggregation}}\\   
$~\bullet~$\textbf{Preparation}

\quad Generate public parameters: $param \leftarrow TS.init(k) $.\\
\quad Each user $i$ in $\mathcal{U}_{0}$:

\quad\quad - Selects random initial secret keys $s_i$, $b_i \in \mathbb{Z}_p$.

\quad\quad - Generates shares: $\{ ( s_i^{j},j ) \}_{j \in \mathcal{U}_{0}} \leftarrow SSS.Share( {s_i} )$ and $\{ ( b_i^{j},j ) \}_{j \in \mathcal{U}_{0}} \leftarrow SSS.Share( {b_i} )$, and distribute them.\\
$~\bullet~$\textbf{Round 0 Advertise Keys}

\quad Server:

\quad\quad - Sends a one-time random value $nonce$ to all users in the user set.

\quad User $i$:

\quad\quad - Computes ${sig \_ s_i} \leftarrow TS.sign( s_i,nonce,param)$ and ${sig \_ b_i} \leftarrow TS.sign(b_i,nonce,param)$. 

\quad\quad - Transforms: $TS.trans(sig \_ s_i)\rightarrow sig \_ sz_i$, $TS.trans(sig \_ b_i)\rightarrow sig \_ bz_i$. $sig \_ sz_i$ and $sig \_ bz_i$ are used as temporary secret 

\quad\quad\quad keys in current aggregation.

\quad There is no need to negotiate communication keys. Other operations remain unchanged.\\
$~\bullet~$\textbf{Round 1 Share Keys}

\quad Omitting this round.\\
$~\bullet~$\textbf{Round 2 Masked Input Collection}

\quad All operations remain unchanged.\\
$~\bullet~$\textbf{Round 3 Unmasking}

\quad User $i$:

\quad\quad - Instead of sending $s_{i}^j$ or $b_{i}^j$, sends ${sig \_ s_{i}^j} \leftarrow TS.sign(s_{i}^j,nonce,param)$ or ${sig \_ b_{i}^j} \leftarrow TS.sign(b_{i}^j,nonce,param)$. 

\quad Server:

\quad\quad - When need to reconstruct $s_i$ or $b_i$, instead of computing $SSS.rec(\cdot)$, computes: $sig \_ s_i \leftarrow TS.rec( {\{ ( sig \_ s_{i}^j, j) \}_{j \in \mathcal{U}_{5}}},t)$

\quad\quad\quad or $sig \_ b_i \leftarrow TS.rec( {\{ ( sig \_ b_{i}^j, j) \}_{j \in \mathcal{U}_{5}}},t)$. 

\quad\quad - Transforms: $TS.trans(sig \_ s_i)\rightarrow sig \_ sz_i$ or $TS.trans(sig \_ b_i)\rightarrow sig \_ bz_i$, which is the temporary secret key the server 

\quad\quad\quad currently requires.

\quad Other operations remain unchanged.\\
\hline
\end{tabular}
\caption{The detailed description of SecAgg + TSKG in the semi-honest setting for one aggregation.}
\label{fig2}
\end{center}
\end{figure*}
\subsection{Proof of Theorem 6.4}
\label{app:5.4}

$Proof.$ Like the proof of Theorem 6.2, we use a standard hybrid argument to prove the theorem. We define a sequence of hybrid distributions ${H}_{0},{H}_{1},{H}_{2}\ldots$ to denote a series of modifications to $REAL$, which can finally get $SIM$. We will prove $SIM$ and $REAL$ are indistinguishable by proving two adjacent hybrids are indistinguishable.
\begin{enumerate}[leftmargin=*]
    \item[$H_0$] In this hybrid, $SIM$ is is exactly the same as $REAL$.
    \item[$H_1$] This hybrid is distributed exactly as the previous one, except all participants register in the PKI at the beginning of the protocol. The distribution of this hybrid is indistinguishable from the previous one.
    \item[$H_2$] This hybrid is distributed exactly as the previous one, except $SIM$ aborts if there is an invalid signature $\sigma_{i}^{1}$. The UF-CMA security of the digital signature scheme can ensure $M_C$ cannot forge any valid signature of an honest user, so the distribution of this hybrid is indistinguishable from the previous one.
    \item[$H_3$] This hybrid is distributed exactly as the previous one, except we replace the communication keys obtained through $DH.agree(\cdot)$ in $\mathcal{U}_1\setminus C$ with uniformly random keys selected by $SIM$. The 2ODH assumption (see Section~\ref{section:kg}) guarantees the distribution of this hybrid is indistinguishable from the previous one.
    \item[$H_4$] This hybrid is distributed exactly as the previous one, except we replace $s_i^j$ sent by $i$ in $\mathcal{U}_2 \setminus C$ to $j$ with a share of $0$. In the unmasking round, honest parties still send the correct shares. Since $| {C \backslash \{ S \}} | < t$, semi-honest parties cannot reconstruct the key, and thus only the plaintext in encryption is changed. The IND-CPA security (as described in Section~\ref{section:ae}) of symmetric authenticated encryption guarantees the distribution of this hybrid is indistinguishable from the previous one.
    \item[$H_5$] This hybrid is distributed exactly as the previous one, except $SIM$ aborts if the message generated by $M_C$ results in a failure of decryption. The IND-CTXT security of symmetric authentication encryption (as described in Section~\ref{section:ae}) guarantees the distribution of this hybrid is indistinguishable from the previous one.
    \item[$H_6$] This hybrid is distributed exactly as the previous one, except $SIM$ aborts if there is an invalid signature $\sigma_{i,j}^{2}$. The UF-CMA security of the digital signature scheme can ensure $M_C$ cannot forge any valid signature of an honest user, so the distribution of this hybrid is indistinguishable from the previous one.
    \item[$H_7$] This hybrid is distributed exactly as the previous one, except we replace the key $s_i$ of $i$ in $\mathcal{U}_3 \setminus C$ with a uniformly random element $s_i'$ in the corresponding field. The randomness of masking keys guarantees the distribution of this hybrid is indistinguishable from the previous one.
    \item[$H_8$] This hybrid is distributed exactly as the previous one, except for the following modification. $M_C$ obtains $\sum_{i \in \mathcal{U}_{3} \cap C}x_{i}$ by calling ${Ideal}_{{\{ x_{i}\}}_{i \in \mathcal{U}\backslash C}}^{\delta}( {\mathcal{U}_{3} \cap C} )$. $SIM$ aborts if there is an illegal request. Users in $\mathcal{U}_3 \setminus C$ instead of sending: 
    \begin{equation}
        \mathbf{y}_{\mathbf{i}} = \left( {x_{i}^{(1)} + E\left( {s_{i}'} \right),x_{i}^{(2)} + E\left( {E\left( {s_{i}'} \right)} \right),\ldots} \right),
    \end{equation}they send: 
    \begin{equation}
        \mathbf{y}_{\mathbf{i}} = \left( {w_{i}^{(1)},w_{i}^{(2)},\ldots} \right),
    \end{equation}where $\{ \mathbf{w}_{\mathbf{i}} \}_{i \in \mathcal{U}_{3} \backslash C}$ are uniformly random vectors satisfying ${\sum_{i \in \mathcal{U}_{3}}\mathbf{w}_{\mathbf{i}}} = {\sum_{i \in \mathcal{U}_{3}}\mathbf{x}_{\mathbf{i}}}$. Lemma~\ref{lemma1} guarantees the distribution of this hybrid is indistinguishable from the previous one.
    \item[$H_{9}$] This hybrid is distributed exactly as the previous one, except $SIM$ aborts if there is an invalid signature $\sigma_{S}^{4}$. The UF-CMA security of the digital signature scheme can ensure $M_C$ cannot forge any valid signature of an honest user, so the distribution of this hybrid is indistinguishable from the previous one.
    \item[$H_{10}$] This hybrid is distributed exactly as the previous one, except $SIM$ aborts if there is an invalid signature $\sigma_{i}^{5}$. The UF-CMA security of the digital signature scheme can ensure $M_C$ cannot forge any valid signature of an honest user, so the distribution of this hybrid is indistinguishable from the previous one.
    \item[$H_{11}$] This hybrid is distributed exactly as the previous one. We have previously discussed that if $t \geq \left\lfloor \frac{2n}{3} \right\rfloor + 1$, then $S$ cannot obtain the sum of secret keys of any user difference set, so the distribution of this hybrid is indistinguishable from the previous one.
    \item[$H_{12}$] This hybrid is distributed exactly as the previous one, except for the following modification. $M_C$ obtains $\sum_{i \in \mathcal{U}_{3} \cap C}s_{i}$ by calling ${Ideal}_{{\{ s_{i}\}}_{i \in \mathcal{U}\backslash C}}^{\delta}( {\mathcal{U}_{3} \cap C} )$. $SIM$ aborts if there is an illegal request. In unmasking round we replace $s_{sum}^{i} = {\sum_{j \in \mathcal{U}_{3}}s_{j}^{i}}$ sent by $i$ in $\mathcal{U}_5 \setminus C$ with ${rand}_{sum}^{i} = {\sum_{j \in \mathcal{U}_{3}}{rand}_{j}^{i}}$, where $rand_j$ is a uniformly random element in the corresponding field satisfying ${\sum_{j \in \mathcal{U}_{3}}{rand}_{j}} = {\sum_{j \in \mathcal{U}_{3}}s_{j}}$, and $rand_j^i$ is a share of $rand_j$. The security of Shamir secret sharing (as described in Section~\ref{section:ss}) guarantees the distribution of this hybrid is indistinguishable from the previous one.
\end{enumerate}

Therefore, the distribution of $SIM$ that has the same distribution as $H_{12}$ is indistinguishable from $REAL$. $SIM$ does not depend on the inputs of honest parties. $C$ can only learn about the sum of input vectors and the sum of secret keys of $\mathcal{U}_{3} \cap C$, where $| {\mathcal{U}_{3} \cap C} | > \delta$. And the above conclusion can still be guaranteed if the protocol aborts in any round. The proof is completed.

\end{document}